\documentclass[%
reprint,preprintnumbers,
nofootinbib,
amsmath,amssymb,
aps
]{revtex4-1}
\pdfoutput=1
\usepackage{graphicx}
\usepackage{soul}
\usepackage[utf8]{inputenc}
\usepackage{flushend,comment}
\usepackage{dcolumn}
\usepackage{bm}
\usepackage{balance}
\usepackage{graphicx}
\usepackage{subcaption}
\usepackage[normalem]{ulem}
\usepackage[colorlinks = true,
            linkcolor = blue,
            urlcolor  = blue,
            citecolor = blue,
            anchorcolor = blue]{hyperref}
\usepackage{verbatim}
\usepackage{color,ulem}
\usepackage[english]{babel}
\usepackage{MnSymbol,wasysym,bbold}
\usepackage[utf8]{inputenc}
\input Starburst.fd
\newcommand*\initfamily{\usefont{U}{Starburst}{xl}{n}}\initfamily 

\newcommand{\beq}{\begin{eqnarray}}
\newcommand{\eeq}{\end{eqnarray}}
\usepackage{amsmath}
\usepackage{tikz}
\usetikzlibrary{decorations.pathmorphing}
\usetikzlibrary{shapes.misc}
\tikzset{cross/.style={cross out, draw=black, minimum size=8*(#1-\pgflinewidth), inner sep=0pt, outer sep=0pt},
cross/.default={1pt}}
\usetikzlibrary{patterns,math}

\newcommand{\Tr}{\operatorname{Tr}}
\newcommand{\ket}[1]{|#1\rangle}
\newcommand{\bra}[1]{\langle #1|}
\newcommand{\avg}[1]{\langle #1\rangle}
\newcommand{\NB}{N_b}

\newcommand{\nn}{\nonumber \\}

\begin{document}

\preprint{\texttt{APCTP Pre2026 - 015}}

\title{Onset of random-matrix statistics in the Yukawa-Sachdev-Ye-Kitaev model:\\ Spectral correlations and Krylov complexity}

\author{Sizheng Cao$^{1}$}\email{caosizheng@fudan.edu.cn}
\author{Hyun-Sik Jeong$^{2,3}$}\email{hyunsik.jeong@apctp.org}
\author{Yi-Li Wang$^{2}$\vspace{0.2cm}}\email{yili.wang@apctp.org}

\affiliation{$^{1}$State Key Laboratory of Molecular Engineering of Polymers, Department of Macromolecular Science, Institute of Fiber Materials and Devices, and Laboratory of Advanced Materials, Fudan University, Shanghai 200438, China}
\affiliation{$^{2}$Asia Pacific Center for Theoretical Physics, Pohang 37673, Korea}
\affiliation{$^{3}$Department of Physics, Pohang University of Science and Technology, Pohang 37673, Korea}

\begin{abstract}
The Yukawa-Sachdev-Ye-Kitaev (YSYK) model, a solvable model of non-Fermi liquids with a proposed cavity-QED realization, couples $N$ complex fermions to $M$ bosons of frequency $\omega_0$ through random Yukawa couplings of strength $g$, and its dynamics depends only on $R\equiv\omega_0/g^{2/3}$. For $R\ll1$ the bosons are slow and act as a static random background in which the fermions are free, while for $R\gg1$ they are only virtually excited and generate a random four-fermion interaction of rank $M$, which is generically chaotic for $M\geq2$. Using exact diagonalization at infinite temperature, we follow the emergence of random-matrix statistics between these limits with unfolded spectral statistics and the Krylov complexity of the thermofield-double state. For complex couplings the level-spacing distribution evolves from Poisson to Gaussian-unitary-ensemble (GUE) statistics, the spectral form factor develops a correlation hole and a ramp, and the Krylov complexity evaluated on the unfolded spectrum develops a peak above its late-time plateau whose height reaches the GUE value. All diagnostics locate the onset around $R\approx0.3$, and adding boson modes moves it to smaller $R$. This delineates the parameter regime in which a cavity-QED realization would display random-matrix spectral correlations. On the other hand, real couplings give the same crossover towards Gaussian-orthogonal-ensemble (GOE) statistics, and along an interpolation between the two the peak height follows the mean gap ratio, so the peak identifies the symmetry class as well as the onset of chaos. Without unfolding the peak stays about 20$\%$ below its random-matrix value in both classes, which we trace to the non-semicircular density of states, and at larger $R$ the boson-number bands slow down the spreading of the state in Krylov space.
\end{abstract}
\maketitle

%
\section{Introduction}\label{sec:intro}
Quantum chaos in many-body systems is usually diagnosed in two ways. The first is spectral. According to the Bohigas-Giannoni-Schmit conjecture~\cite{Bohigas:1983er,Bohigas}, the energy levels of a chaotic system repel each other and carry the correlations of random matrices~\cite{Meh2004,GUHR1998189}, while the levels of a generic integrable system are uncorrelated~\cite{Berry_1977}. The distribution of nearest-neighbour spacings tests these correlations on the scale of one mean level spacing. The spectral form factor (SFF) tests them on all scales; its ramp, the signature of long-range spectral rigidity, is resolved at times comparable to the Heisenberg time. The second way is dynamical: the early-time growth of out-of-time-order correlators (OTOCs)~\cite{larkin1969quasiclassical} defines a quantum Lyapunov exponent, which is bounded by $2\pi k_B T/\hbar$~\cite{Maldacena2016} for the maximally chaotic systems. Both are tied to chaos, thermalization and to the scrambling of quantum information~\cite{DAlessio:2016aa,Haake_2018}. Nevertheless, for systems with a few degrees of freedom, where neither large-$N$ nor semiclassical methods apply, one would like a dynamical quantity that can be computed exactly and that carries the information contained in the spectral statistics.

The Sachdev-Ye-Kitaev (SYK) model~\cite{Sachdev:1992fk,Kitaev2015Talk,Chowdhury:2021qpy,Rosenhaus:2018dtp} of $N$ fermions with random all-to-all $q$-body interactions has become the standard setting for such questions. It is solvable at large $N$, develops an emergent conformal symmetry at low energies with a dual description in nearly-AdS$_2$ Jackiw-Teitelboim gravity~\cite{Maldacena:2016hyu,Kitaev:2017awl}, and for $q\geq4$ saturates the chaos bound at low temperature ($T$). The case $q=2$ (SYK$_2$) is different in kind. The Hamiltonian is then quadratic, so the theory can be a free (integrable) fermion system whose many-body energies are sums of $N/2$ independent numbers and consequently do not repel each other. For $q=4$ (SYK$_4$) the many-body spectrum itself follows random-matrix theory, and the SFF develops a ramp structure before the Heisenberg time~\cite{Cotler:2016fpe}. Adding a quadratic term to the quartic model interpolates between the two cases and drives a chaotic-integrable transition~\cite{Garcia-Garcia:2017bkg}.

The Yukawa-SYK (YSYK) model~\cite{Esterlis:2019ola,Wang:2020aa,Wang:2020dtj,Pan:2020mqy} offers a tractable approach to the chaotic-integrable transition through a physically motivated different mechanism. In the YSYK model the fermions have no direct interaction. Instead, $N$ complex fermions couple to $M$ bosons of frequency $\omega_0$ through random Yukawa couplings of variance $g^2$. The model was introduced to describe electrons coupled to soft bosonic modes and has been used to study non-Fermi liquids, pairing and self-tuned criticality~\cite{Esterlis:2019ola,Wang:2020aa,Wang:2020dtj,Pan:2020mqy,Choi:2021hvi,HAUCK2020168120,Kim:2020jpz}. Its spatially extended version provides a theory of strange metals~\cite{Patel:2022gdh}, and its pairing problem has been related to holographic superconductivity in AdS$_2$~\cite{Stangier:2022nvq,Stangier:2026ija}. From the point of view of quantum chaos~\cite{PascualSolis:2025cuc,Davis:2022iqi,Marcus:2018tsr,Kim:2019lwh}, the model has a single dimensionless parameter, $R\equiv\omega_0/g^{2/3}$, which interpolates between the two SYK limits. At small $R$ the bosons are slow and the fermions move in an essentially static random field, which leaves them effectively non-interacting. At large $R$ the bosons are virtually excited, and their exchange produces a four-fermion interaction. A finite-size study based on the density of states, the gap ratio, the SFF and OTOCs~\cite{PascualSolis:2025cuc} established this crossover from single-particle to many-body chaos and proposed a cavity-QED realization in which cavity photons mediate the Yukawa coupling. At large $N$, ladder resummations of the OTOC~\cite{Davis:2022iqi} give a sizable Lyapunov exponent in the non-Fermi-liquid phase and an exponentially suppressed one in the gapped insulating phase, and closely related boson-fermion models and low-rank SYK models can be maximally chaotic~\cite{Marcus:2018tsr,Kim:2019lwh}.

Krylov, or spread, complexity~\cite{Balasubramanian:2022tpr,Caputa:2024vrn} (see~\cite{Nandy:2024evd,Rabinovici:2025otw,Jeong:2026gdc} for reviews) is a complementary dynamical probe of quantum chaos that needs only a Hamiltonian and a reference state.\footnote{Nevertheless, unlike the OTOC, Krylov complexity has also been confirmed to be a robust and universal probe of chaos beyond large-$N$ systems, including one- or few-body systems~\cite{Nandy:2024evd,Rabinovici:2025otw,Jeong:2026gdc}.} Starting from the reference state, the Lanczos algorithm~\cite{Lanczos:1950zz} builds an orthonormal basis in which the Hamiltonian is tridiagonal. The evolution then becomes that of a particle hopping on a half-line, and the Krylov complexity is the mean position of this particle. The Krylov basis minimizes the spread of the evolving state among ordered bases, at least at early times~\cite{Balasubramanian:2022tpr}. When the reference state is the infinite-temperature thermofield-double (TFD) state, the survival probability equals the SFF, and the Krylov complexity becomes a functional of the spectrum alone. At early times it grows as $b_1^2t^2$, with $b_1$ the first Lanczos coefficient~\cite{Erdmenger:2023wjg,Huh:2023jxt}. In many chaotic systems it then rises above its late-time saturation value, reaches a peak and relaxes to a plateau fixed by the dimension of the Krylov space, whereas in integrable systems it approaches the plateau without a peak~\cite{Balasubramanian:2022tpr}. The height of the peak has been used as an order parameter for chaotic-integrable transitions~\cite{Baggioli:2024wbz,Erdmenger:2026iga}, mostly in two deformations of the Majorana SYK model~\cite{Balasubramanian:2022tpr,Baggioli:2024wbz,Huh:2024ytz}: the mass-deformed model, in which a random quadratic term competes with the quartic one~\cite{Garcia-Garcia:2017bkg,Huh:2024ytz}, and the sparse model, in which a fraction of the quartic couplings is removed~\cite{Baggioli:2024wbz}. The peak is not fixed by level correlations alone, it also depends on the smooth density of states, which can raise or lower the peak for reasons that have nothing to do with chaos, for instance the saddle-dominated scrambling phenomena~\cite{Xu:2019lhc,Huh:2023jxt}. Evaluating Krylov complexity on the unfolded spectrum removes this dependence~\cite{Erdmenger:2026iga,Caputa:2026hvh,Basu:2026gvl}.

In the mass-deformed and sparse models the transition is engineered at the level of a fixed quartic vertex. The YSYK model reaches the same two limits ($q=2$ or $q=4$ SYK-like limits) through a different mechanism: the interaction is mediated by a separate dynamical field, and its effective form changes with $R$. The model is also a demanding test for the Krylov complexity, because its density of states changes qualitatively along the crossover~\cite{PascualSolis:2025cuc}. It is smooth and close to Gaussian at small $R$, and at large $R$ it splits into bands labelled by the total boson number. Krylov complexity has not been applied to this boson-mediated crossover in the literature yet, and the spectral statistics of the model have not yet been analysed after unfolding.\footnote{Ref.~\cite{PascualSolis:2025cuc} studied spectral statistics such as the SFF without unfolding the spectrum, whereas the gap ratio does not require unfolding at all, and explored the Poisson-to-GUE crossover.}

In this paper we study the finite-size spinless YSYK model to systematically probe the onset of random-matrix level statistics. We first derive the two limits in Hamiltonian form (section~\ref{sec:model}). For $R\ll1$ the boson coordinates become conserved and the model splits into independent free-fermion problems. For $R\gg1$ a Schrieffer-Wolff transformation gives, in each band of fixed boson number, a four-fermion interaction of rank $M$, which is integrable for $M=1$ and generically chaotic for $M\geq2$. We also identify the antiunitary symmetries of the model: there are none for complex couplings and one, squaring to $+1$, for real couplings, so the chaotic regime belongs to the Gaussian unitary (GUE) and orthogonal (GOE) ensemble, respectively. We then follow the crossover with three diagnostics, the unfolded spacing distribution, the unfolded SFF, and the Krylov complexity computed from the raw and from the unfolded spectrum, for complex couplings, for real couplings, and along an interpolation between them. Our main findings are the following. (i) All diagnostics place the onset of random-matrix statistics at $R\approx0.3$, and more boson modes shift it to smaller $R$. (ii) The peak of the unfolded Krylov complexity reaches the value of the random-matrix theories of Gaussian ensemble, and along the GOE-GUE interpolation it follows the mean gap ratio; its height therefore identifies the symmetry class as well as the onset of chaos. (iii) The raw peak, from Krylov complexity with raw spectrum, remains below its random-matrix value.

The paper is organized as follows. Section~\ref{sec:model} introduces the model, its symmetries, and its two limits. Section~\ref{sec:diagnostics} defines the diagnostics of quantum chaos. Section~\ref{sec:spectral} presents the spectral statistics, while Section~\ref{sec:krylov} presents the Krylov complexity and the Lanczos coefficients. We conclude in Section~\ref{sec:discussion} with a summary of results and perspectives for future directions.

\begin{figure*}[t]
\centering
\includegraphics[scale=0.3]{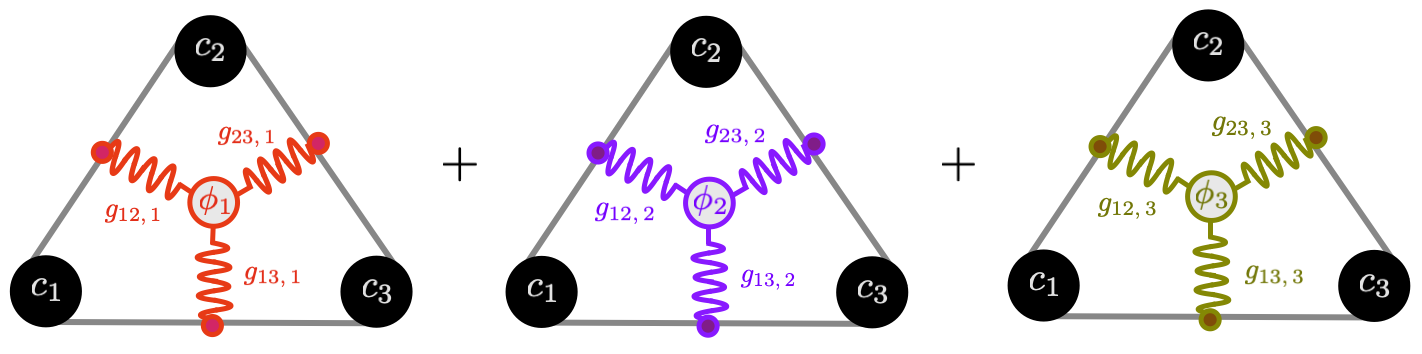}

\vspace{20pt}

\includegraphics[scale=0.35]{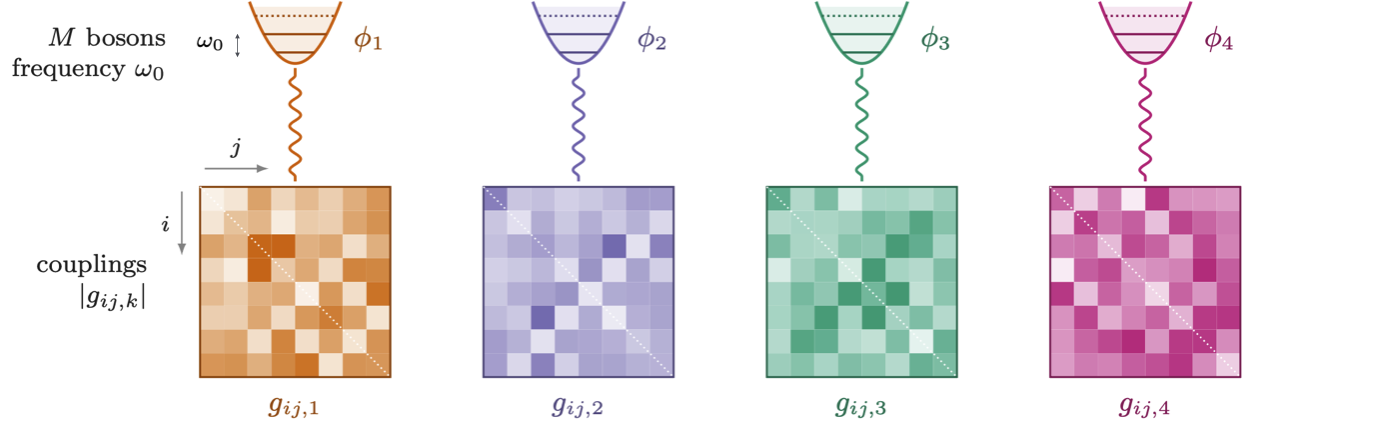}
\caption{Structure of the Yukawa-SYK model. Top: for a toy example of $N=3$ fermions and $M=3$ bosons, each boson $\phi_k$ mediates a random coupling $g_{ij,k}$ between every pair of fermion bilinears $c_i^\dagger c_j$ (diagonal terms $g_{ii,k}c_i^\dagger c_i$ are not shown); the full Hamiltonian sums this vertex over $k=1,\dots,3$. Bottom: for the sizes used in the numerics below ($N=8$, $M=4$), the couplings $g_{ij,k}$ for each boson form an independent $N\times N$ Hermitian random matrix (GUE for complex, GOE for real couplings), shown here by $|g_{ij,k}|$ for one disorder realization.}
\label{fig:model}
\end{figure*}
%

%
\section{Model and limiting regimes}\label{sec:model}

\subsection{Hamiltonian, symmetries and truncation}\label{sec:hamiltonian}

Let us first revisit the definition and basic properties of the Yukawa-SYK (YSYK) model~~\cite{Esterlis:2019ola,Wang:2020aa,Wang:2020dtj,Pan:2020mqy}. We consider the ($0+1$)-dimensional YSYK model of $N$ spinless complex fermions $c_i$ and $M$ real bosons $\phi_k$ with conjugate momenta $\pi_k$ at zero chemical potential,
\begin{align}\label{eq:H}
\begin{split}
H&=\frac12\sum_{k=1}^{M}\left(\pi_k^2+\omega_0^2\phi_k^2\right)+\frac{1}{\sqrt{MN}}\sum_{i,j=1}^{N}\sum_{k=1}^{M}g_{ij,k}\,c_i^\dagger c_j\,\phi_k,
\end{split}
\end{align}
with Gaussian random couplings satisfying
\begin{equation}
g_{ij,k}=g_{ji,k}^{*},\qquad \avg{g_{ij,k}}=0,\qquad \avg{|g_{ij,k}|^2}=g^2 .
\label{eq:couplings}
\end{equation}
Figure~\ref{fig:model} illustrates the schematic structure of YSYK model: each boson mediates a random coupling between every pair of fermions, and for fixed $k$ the couplings $g_{ij,k}$ form an $N\times N$ Hermitian random matrix.
Unless stated otherwise the couplings are complex, so that for each $k$ the matrix $g_k$ with entries $g_{ij,k}$ is drawn from the GUE. Real couplings (GOE) and the interpolation between the two cases are introduced in section~\ref{sec:alpha}.

The Hamiltonian is invariant under the U(1) transformation $c_i\to e^{i\theta}c_i$, and the fermion number $Q=\sum_ic_i^\dagger c_i$ is conserved. We work in the half-filled sector $Q=N/2$, where the chemical potential only shifts the energy and can be dropped. In terms of $\phi_k=(a_k+a_k^\dagger)/\sqrt{2\omega_0}$ and $\pi_k=i\sqrt{\omega_0/2}\,(a_k^\dagger-a_k)$, the Hamiltonian \eqref{eq:H} can be rewritten
\begin{equation}
H=\omega_0\sum_{k=1}^{M}\Big(a_k^\dagger a_k+\frac12\Big)+\lambda\sum_{k=1}^{M}\rho_k\,x_k,
\label{eq:Hosc}
\end{equation}
where
\begin{equation}
\rho_k\equiv\sum_{i,j=1}^Ng_{ij,k}\,c_i^\dagger c_j,\quad x_k\equiv a_k+a_k^\dagger,\quad \lambda\equiv\frac{1}{\sqrt{2\omega_0MN}}.
\end{equation}
The operators $\rho_k$ are Hermitian fermion bilinears, one for each boson mode.

We keep the occupations for bosons $n_k=0,\dots,\NB$ and define the model by eq.~\eqref{eq:Hosc} with $a_k$ restricted to this space. The half-filled sector then has dimension
\begin{equation}
D=\binom{N}{N/2}(\NB+1)^M .
\label{eq:D}
\end{equation}
All numerical results use $\NB=1$, for which each boson becomes a two-level system, $x_k\to\sigma^x_k$ and $a_k^\dagger a_k\to(1-\sigma_k^z)/2$. $\sigma^x_k, \sigma^z_k$ are Pauli matrices acting on the two-state space $\{\ket{0}_k, \ket{1}_k\}$ of boson $k$, with $\sigma^z_k\ket{0}_k=\ket{0}_k$ and $\sigma^z_k\ket{1}_k=-\ket{1}_k$. For $N=8$ and $M=4$, for instance, the Hilbert-space dimension gives $D=1120$.
The fermions are represented by a Jordan-Wigner transformation, $c_i=\prod_{j<i}(1-2\hat{n}_j)\,\tau_i^-$ with $\hat{n}_j=c_j^\dagger c_j$ and $\tau^-=\ket{0}\bra{1}$ with $\ket{0}$ and $\ket{1}$ denote the states of single fermionic mode. In this basis the matrices of $c_i$, $c_i^\dagger$, $a_k$ and $a_k^\dagger$ are real, a fact used in section~\ref{sec:symmetry}. For each realization of the couplings we build the Hamiltonian in the half-filled sector and diagonalize it exactly.

The rescaling $\omega_0\to\Lambda\omega_0$, $g\to\Lambda^{3/2}g$ multiplies eq.~\eqref{eq:Hosc} by $\Lambda$,
\begin{equation}
H(\Lambda\omega_0,\Lambda^{3/2}g)=\Lambda\,H(\omega_0,g).
\label{eq:scaling}
\end{equation}
The spectrum in units of $g^{2/3}$, and every quantity built from it, depends on $\omega_0$ and $g$ only through $R=\omega_0/g^{2/3}$, while time enters only through $g^{2/3}t$. For the Krylov complexity defined below, for example, $C(t;\omega_0,g)=C(g^{2/3}t;R)$. We set $g=1$ in all numerical results, so that $\omega_0=R$; the figures label the frequency axis by $\omega_0$.

The meaning of $R$ becomes clear from the two energy scales of eq.~\eqref{eq:Hosc}. The Yukawa term has the scale $E_Y\equiv g/\sqrt{\omega_0}$; for a frozen boson configuration this is, up to a factor of order one, the single-particle bandwidth of the fermions (the $1/\sqrt{MN}$ normalization makes it independent of $N$ and $M$): see the discussion below eq.~\eqref{eq:hvar}. The interaction induced by boson exchange has the scale $E_Y^2/\omega_0=g^2/\omega_0^2$. Hence
\begin{equation}
R^{3/2}=\frac{\omega_0}{E_Y} \quad \text{or} \quad R^{3}=\frac{\omega_0}{g^2/\omega_0^2}.
\label{eq:scales}
\end{equation}
Equivalently, $R^{3/2}$ is the ratio of the boson frequency to the fermionic bandwidth, and $R^3$ is the ratio of the boson frequency to the scale of the boson-induced interaction. Therefore, small $R$ means bosons that are slow compared with the fermions; large $R$ means bosons that are fast and an induced interaction that is weak compared with $\omega_0$. At fixed $\omega_0$ a larger $g$ makes the bosons slower relative to the fermions, driving the system toward the free-fermion regime of section~\ref{sec:smallR}, which, as we now show, makes the spectrum less chaotic.

\subsection{\texorpdfstring{$R\ll1$}{R << 1}: static bosons and free fermions}\label{sec:smallR}

For $\omega_0\ll E_Y$ the oscillator term in eq.~\eqref{eq:Hosc} is a small perturbation of
\begin{equation}
H_Y=\lambda\sum_{k=1}^M\rho_k\,x_k .
\label{eq:HY}
\end{equation}
The operators $x_k$ commute with each other and with all fermion operators, so they are conserved by $H_Y$. In their joint eigenbasis, $x_k\ket{\xi}=\xi_k\ket{\xi}$, where the $\xi_k$ are the zeros of the Hermite polynomial ($\xi_k=\pm1$ for $\NB=1$), the Hamiltonian becomes block diagonal,
\begin{equation}
H_Y=\bigoplus_{\xi}\;\sum_{i,j=1}^N h_{ij}(\xi)\,c_i^\dagger c_j,\qquad h(\xi)=\lambda\sum_{k=1}^M\xi_k\,g_k .
\label{eq:blocks}
\end{equation}
Each block is a free-fermion, SYK$_2$-like Hamiltonian. Its single-particle matrix $h(\xi)$ is a Gaussian random matrix of the same symmetry class as the $g_k$, so the single-particle levels have random-matrix statistics~\cite{PascualSolis:2025cuc}. The many-body spectrum, however, is a superposition of $(\NB+1)^M$ free-fermion spectra. Within each block the many-body energies are sums of $N/2$ single-particle energies, and levels in different blocks belong to different sectors of the conserved $x_k$. Neither mechanism produces level repulsion, and we expect Poisson statistics for real and for complex couplings alike.

It is worth noting that the resulting single-particle Hamiltonian $h_{ij}(\xi)=\lambda\sum_k\xi_kg_{ij,k}$ can be an $N\times N$ random Hermitian matrix whose entries have variance
\begin{equation}\label{eq:hvar}
\mathrm{Var}[h_{ij}]=\lambda^2\sum_{k=1}^M\xi_k^2\,g^2=\frac{g^2}{2\omega_0N},
\end{equation}
using $\xi_k^2=1$ for $\NB=1$ and $\lambda^2=1/(2\omega_0MN)$; the sum over $M$ independent channels cancels the explicit $M$ in $\lambda^2$. By the semicircle law, an $N\times N$ random matrix with entry variance $\sigma^2$ has eigenvalues spread over a range $\sim\sigma\sqrt N$; here this gives $\sqrt{N\cdot g^2/(2\omega_0N)}=g/\sqrt{2\omega_0}\sim E_Y$, independent of $N$ as well. The normalization $1/\sqrt{MN}$ in eq.~\eqref{eq:H} is fixed precisely by these two cancellations, so that $E_Y$ is a genuine single-particle energy scale of the frozen-boson problem rather than an artifact of the counting of fermions or boson modes.

Equation~\eqref{eq:blocks} is the operator form of the static limit of the boson propagator. One can find that integrating out $\phi_k$ in the Euclidean path integral produces the kernel $(\nu^2+\omega_0^2)^{-1}$ between fermion bilinears; for $\omega_0\to0$ it is dominated by $\nu=0$, so that the fermions couple only to the time-independent part of each $\phi_k$, which can be traded for one static Hubbard-Stratonovich variable per boson mode.\footnote{We provide an alternative path-integral interpretation of the SYK$_2$-to-SYK$_4$ crossover in Appendix \ref{app:pi}.}

The leading correction is the oscillator term itself. For $\NB=1$ it equals $-(\omega_0/2)\sum_k\sigma_k^z$ up to a constant, flips $\xi_k\to-\xi_k$, and connects every block to $M$ others whose single-particle Hamiltonians differ by $2\lambda\xi_kg_k$, a change of order $E_Y$. It acts like a transverse field that tunnels between blocks with amplitude $\omega_0/2$, and its strength relative to the fermionic dynamics is $R^{3/2}$. Level repulsion requires these processes to hybridize many-body levels of different blocks. This should happen first for nearly degenerate levels and then spread to larger energy separations as $R$ grows. Both the number of neighbouring blocks, $M$, and the reduction of the level spacing with growing $D$ favour an earlier onset for larger $M$.

\begin{figure*}[t]
\centering
\includegraphics[width=\linewidth]{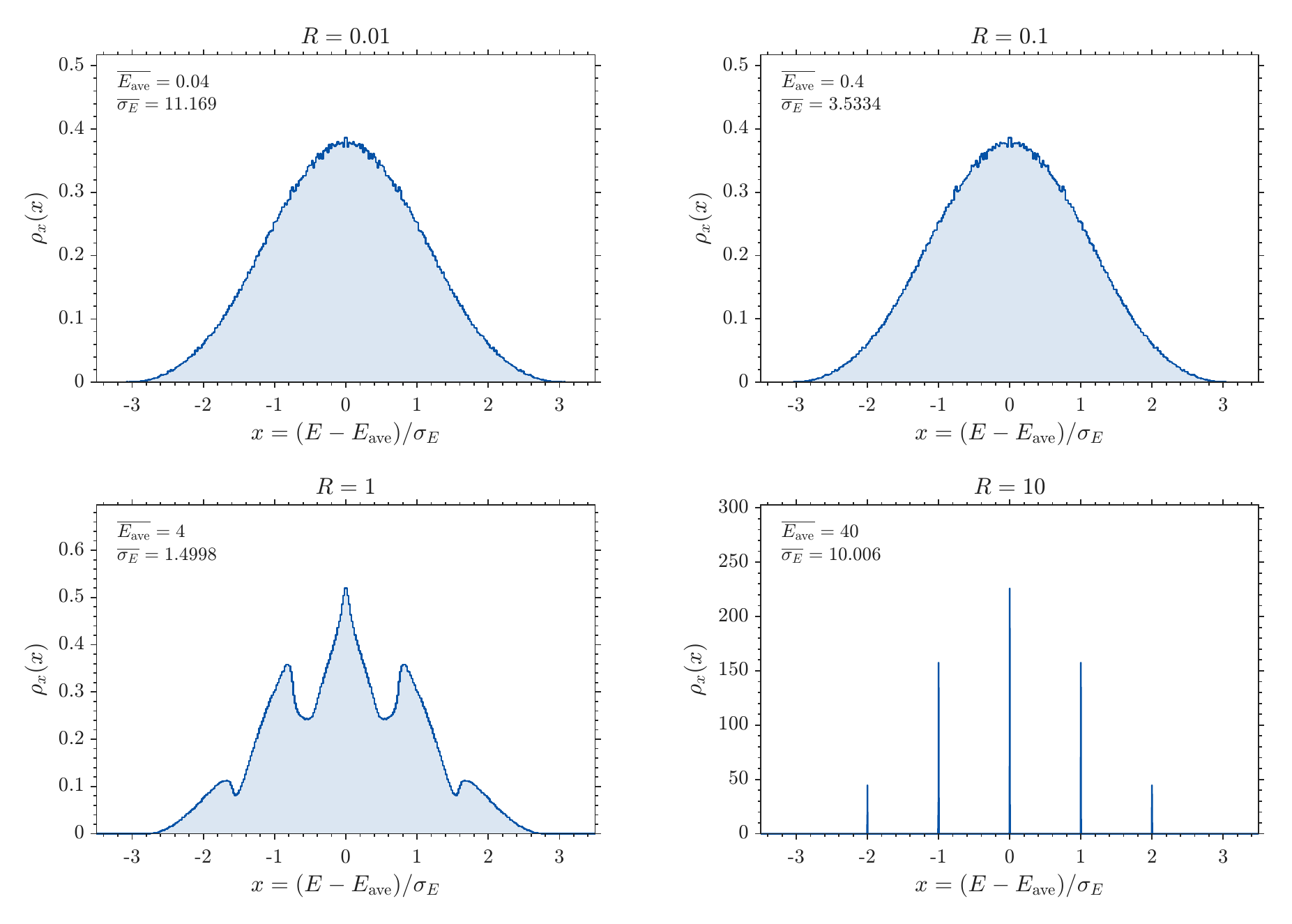}
\caption{Disorder-averaged density of states of the YSYK model at $R=0.01,0.1,1,10$ (complex couplings, $N=8$, $M=4$, $\NB=1$, $1000$ realizations), plotted against $x=(E-\overline{E_{\rm ave}})/\overline{\sigma_E}$, where $E_{\mathrm{ave}}$ and $\sigma_E$ are its spectral mean and standard deviation. At small $R$ the density is smooth, the superposition of the free-fermion spectra of section~\ref{sec:smallR}. By $R=1$ five lobes are already visible, and at $R=10$ the spectrum has split into the $M+1=5$ bands of section~\ref{sec:largeR}, with relative populations close to the predicted $70:280:420:280:70$.}
\label{fig:DOS}
\end{figure*}

\subsection{\texorpdfstring{$R\gg1$}{R >> 1}: virtual bosons and a rank-\texorpdfstring{$M$}{M} interaction}\label{sec:largeR}

For $E_Y\ll\omega_0$ the Yukawa term only virtually excites the bosons and can be removed perturbatively. The Schrieffer-Wolff generator $S=(\lambda/\omega_0)\sum_k\rho_k(a_k^\dagger-a_k)$ satisfies $[S,\omega_0\sum_ka_k^\dagger a_k]=-\lambda\sum_k\rho_kx_k$, and $e^{S}He^{-S}$ gives, to second order in $\lambda$,
\begin{align}\label{eq:SW}
\begin{split}
H_{\rm eff}&=\omega_0\sum_{k}\Big(a_k^\dagger a_k+\frac12\Big)-\frac{1}{2\omega_0^2MN}\sum_{k}\rho_k^2 \\
 &\qquad+ \frac{1}{2\omega_0^2MN}\sum_{k<k'}\big[\rho_k,\rho_{k'}\big]\big(a_k^\dagger a_{k'}-a_{k'}^\dagger a_k\big).
\end{split}
\end{align}
We have dropped terms that change the boson number by two, which connect states $2\omega_0$ apart and contribute only at higher order, together with corrections suppressed by further powers of $R^{-3/2}$. The effective Hamiltonian conserves the total boson number $\sum_ka_k^\dagger a_k$, whose eigenvalues we denote by $\ell$. The spectrum consists of $M\NB+1$ bands centred near $\omega_0(\ell+M/2)$, each with a width proportional to $g^2/\omega_0^2=\omega_0R^{-3}$, while the Yukawa term, which couples neighbouring bands, is of relative order $R^{-3/2}$ compared with their spacing. For $\NB=1$ and $M=4$ the five bands contain $70\binom{4}{\ell}=70,280,420,280,70$ states: see figure \ref{fig:DOS} for the density of states of YSYK model.

In the lowest band, $\ell=0$, the last term of eq.~\eqref{eq:SW} vanishes, and
\begin{align}\label{eq:lowrank}
\begin{split}
H_{\rm eff}^{(0)}&=-\frac{1}{2\omega_0^2MN}\sum_{k=1}^M\rho_k^2 \\
&=-\frac{1}{2\omega_0^2N}\sum_{i,j,i',j'=1}^{N}J_{ij;i'j'}\,c_i^\dagger c_j\,c_{i'}^\dagger c_{j'},
\end{split}
\end{align}
where
\begin{equation}
\begin{gathered}
J_{ij;i'j'}=\frac1M\sum_{k=1}^Mg_{ij,k}\,g_{i'j',k}.
\end{gathered}
\end{equation}
Since virtual processes out of the boson vacuum involve only single quanta, this result is exact at this order for any $\NB\geq1$. For $\NB=1$ the effective Hamiltonian in the other bands has the same structure, except that the $\rho_k^2$ term changes sign for occupied modes: an excited two-level system can only be virtually de-excited.

Viewed as a matrix acting on the $N^2$-dimensional space of index pairs $(ij)$, the coupling $J$ has rank at most $M$. This structure explains the role of $M$ at large $R$. For $M=1$, $H_{\rm eff}^{(0)}\propto-\rho_1^2$ is a function of a single quadratic operator; its eigenstates are Slater determinants built from the eigenvectors of $g_1$, and the spectrum is integrable. For $M\geq2$ the $\rho_k$ do not commute for generic couplings, and eq.~\eqref{eq:lowrank} is a genuine many-body interaction of the low-rank SYK type~\cite{Kim:2019lwh}. Random-matrix statistics do not require Gaussian couplings, so we expect them for any $M\geq2$, with larger $M$ bringing the interaction closer to a generic complex SYK$_4$ model. In the limit $M\to\infty$ this becomes exact. By the central limit theorem, $J=\avg{J}+\delta J$, where the mean $\avg{J_{ij;i'j'}}=G_{ij;i'j'}\equiv\avg{g_{ij,k}g_{i'j',k}}$ equals $g^2\delta_{ij'}\delta_{ji'}$ for complex couplings (with an extra term $g^2\delta_{ii'}\delta_{jj'}$ for real ones). The fluctuation $\delta J$ becomes Gaussian with
\begin{align}\label{eq:covariance}
\begin{split}
\avg{\delta J_{ij;i'j'}\,\delta J_{mn;m'n'}}&=\frac1M\Big(G_{ij;mn}\,G_{i'j';m'n'} \\ &\qquad\quad + \, G_{ij;m'n'}\,G_{i'j';mn}\Big),
\end{split}
\end{align}
which is the covariance of a complex SYK$_4$ coupling with variance of order $g^4/M$. Normal ordering the quartic operator also produces a random one-body term, whose size relative to the two-body term is of order $N^{-1/2}$. The large-$R$, large-$M$ limit of the lowest band is thus a complex SYK$_4$ model with extra terms containing only fermion numbers.

\subsection{Antiunitary symmetry and the Dyson class}\label{sec:symmetry}
The random-matrix class of the chaotic regime is fixed by the antiunitary symmetries that commute with $H$ within a fixed-charge sector~\cite{Haake_2018,dyson1962statistical,dyson1962threefold}. Since the matrices of $c_i$, $c_i^\dagger$, $a_k$ and $a_k^\dagger$ are real in the occupation basis, complex conjugation $K$ in this basis acts as $KH(g)K=H(g^*)$. For real couplings $T=K$ is an antiunitary symmetry with $T^2=1$, and the chaotic spectrum should follow the GOE. For complex couplings $K$ is broken.

At half filling there is one more candidate. Particle-hole conjugation $C=\prod_i(c_i+c_i^\dagger)$ maps $c_i\to\pm c_i^\dagger$ (with an $i$-independent sign) and $Q\to N-Q$, and hence maps the half-filled sector onto itself. Combined with boson parity $P_B:a_k\to-a_k$, and using $c_i^\dagger c_j\to\delta_{ij}-c_j^\dagger c_i$ together with $g_{ji,k}=g_{ij,k}^*$, it acts as
\begin{equation}
(CP_B)\,H(g)\,(CP_B)^{-1}=H(g^*)-\lambda\sum_{k=1}^M(\Tr g_k)\,x_k .
\label{eq:PH}
\end{equation}
If all traces $\Tr g_k=\sum_ig_{ii,k}$ vanished, $KCP_B$ would be an antiunitary symmetry for complex couplings and $CP_B$ a unitary one for real couplings. For the vertex in eq.~\eqref{eq:H} the traces are Gaussian variables of variance of order $Ng^2$ and break both. (For a particle-hole symmetric vertex $g_{ij,k}(c_i^\dagger c_j-\delta_{ij}/2)\phi_k$ the half-filled sector would acquire an additional symmetry, antiunitary for complex and unitary for real couplings, similar to the particle-hole symmetry of the complex SYK model~\cite{Gu:2019jub}.) The half-filled sector thus has no antiunitary symmetry for complex couplings and a single one with $T^2=1$ for real couplings, and the chaotic regime should follow the GUE and the GOE, respectively. At large $R$ the same conclusion follows directly from eq.~\eqref{eq:lowrank}, whose coupling $J$ is real exactly when the $g_k$ are.

%
\section{Diagnostics}\label{sec:diagnostics}

\subsection{Unfolding and spectral statistics}\label{sec:unfolding}

In order to study the quantum chaos properties of the spectrum, one has to separate genuine level correlations from the smooth variation of the mean level density, which figure~\ref{fig:DOS} shows to be far from semicircular and, at large $R$, organized into bands. This separation is achieved by unfolding: for each realization we order the levels, $E_1\leq\dots\leq E_D$, fit the level staircase with a smooth function $\mathcal{N}(E)$, and define unfolded levels $\varepsilon_n=\mathcal{N}(E_n)$, rescaled to unit mean spacing. For the spacing distribution and the SFF we keep the central 80\% of the levels ($D_{80}=896$ for $D=1120$) and discard the edges, where the density is small and the fit least reliable.

The distribution $P(s)$ of the unfolded spacings $s_n=\varepsilon_{n+1}-\varepsilon_n$ is compared with the Poisson distribution $P_{\rm P}(s)=e^{-s}$ and with the Wigner surmises $P_{\rm GOE}(s)=\frac{\pi}{2}se^{-\pi s^2/4}$ and $P_{\rm GUE}(s)=\frac{32}{\pi^2}s^2e^{-4s^2/\pi}$, whose linear and quadratic behaviour at small $s$ distinguishes the two classes. The unfolded spectral form factor (SFF) is
\begin{equation}
K(t)=\Big\langle\Big|\sum_{n=1}^{D_{80}}e^{-i\varepsilon_nt}\Big|^2\Big\rangle ,
\label{eq:SFF}
\end{equation}
where $t$ is measured in units of the inverse mean level spacing, so that the Heisenberg time is $t_H=2\pi$. $K(t)$ starts at $D_{80}^2$ and, for a non-degenerate spectrum, fluctuates around the plateau $D_{80}$ for $t>t_H$. For uncorrelated levels it reaches the plateau once the disconnected part has decayed. For correlated levels it first drops below the plateau (the correlation hole) and then rises along a ramp. The sharp edges of the 80\% window add oscillations at early times, which we do not interpret.

As a complementary measure of local correlations we use the ratio of consecutive raw spacings, $r_n=\min(s_n,s_{n+1})/\max(s_n,s_{n+1})$ with $s_n=E_{n+1}-E_n$, averaged over the spectrum and over realizations. The local density cancels in the ratio, so no unfolding is needed. The reference values are $\avg{r}_{\rm P}=2\ln2-1\simeq0.386$, $\avg{r}_{\rm GOE}\simeq0.5307$ and $\avg{r}_{\rm GUE}\simeq0.5996$~\cite{Atas2013}.

\begin{figure*}[t]
\centering
{\small\itshape Complex couplings}\\[1pt]
\includegraphics[scale=0.86]{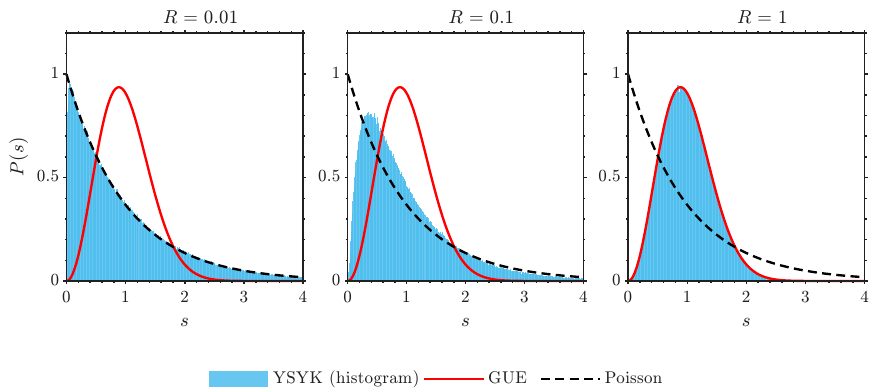}\\[6pt]

\vspace{10pt}

{\small\itshape Real couplings}\\[1pt]
\includegraphics[scale=0.23]{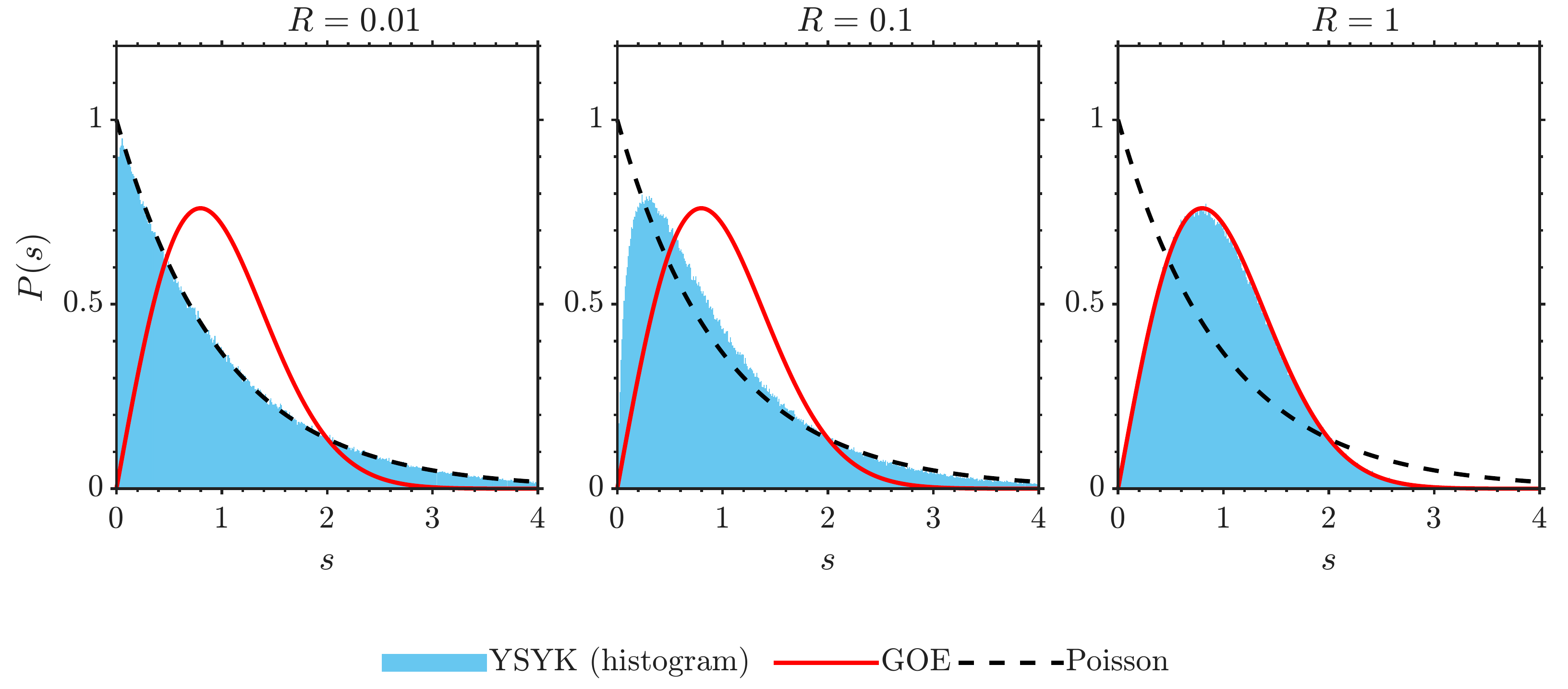}
\caption{Nearest-neighbour spacing distributions at $R=0.01$, $0.1$ and $1$ (the panel titles give $\omega_0$ in units of $g^{2/3}$, i.e.\ $R$), for complex (upper row) and real (lower row) couplings. Histograms: YSYK model with $N=8$, $M=4$, $\NB=1$, and $2000$ realizations. Dashed: Poisson distribution. Solid: Wigner surmise of the GUE and GOE.}
\label{fig:Ps}
\end{figure*}

\begin{figure*}[t]
\centering
{\small\itshape Complex couplings}\\[1pt]
\includegraphics[scale=1.0]{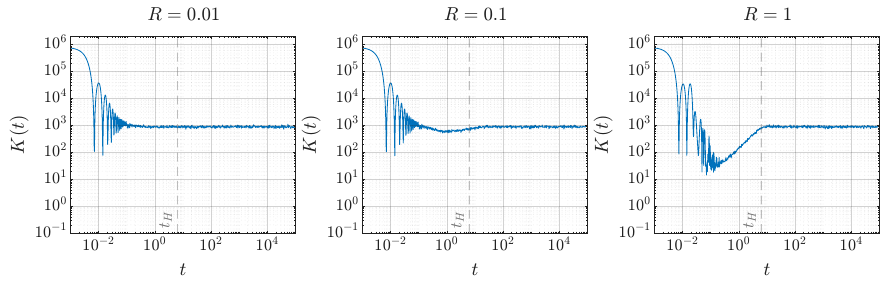}\\[6pt]

\vspace{10pt}

{\small\itshape Real couplings}\\[1pt]
\includegraphics[scale=1.0]{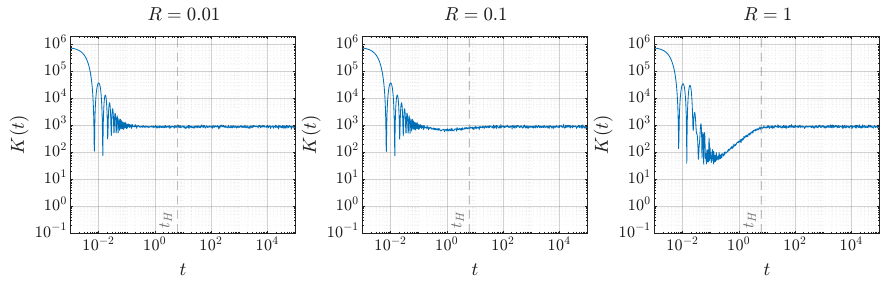}
\caption{Unfolded spectral form factor $K(t)$, eq.~\eqref{eq:SFF}, for $R=0.01$, $0.1$ and $1$ (left to right) with complex (upper row) and real (lower row) couplings; $N=8$, $M=4$, $\NB=1$, and $500$ realizations. All panels share the same axes. The Heisenberg time $t_H=2\pi$ is marked on the horizontal axis.}
\label{fig:SFF}
\end{figure*}

\subsection{Krylov complexity of the thermofield double}\label{sec:spread}

Starting from a normalized reference state $\ket{K_0}$, the Lanczos recursion
\begin{equation}
H\ket{K_n}=a_n\ket{K_n}+b_{n+1}\ket{K_{n+1}}+b_n\ket{K_{n-1}},\qquad b_0=0,
\label{eq:lanczos}
\end{equation}
generates an orthonormal basis $\{\ket{K_n}\}_{n=0}^{D_K-1}$ that terminates when $b_{D_K}=0$ with the Krylov subspace dimension $D_K\leq D$. In this basis the Hamiltonian is tridiagonal, and the evolution is that of a particle on a chain with on-site energies $a_n$ and hoppings $b_n$~\cite{Viswanath_1994}. Writing $\ket{\psi(t)}=e^{-iHt}\ket{K_0}=\sum_n\varphi_n(t)\ket{K_n}$, the Krylov, or spread, complexity~\cite{Balasubramanian:2022tpr} is the mean position on the chain,
\begin{equation}
C(t)=\sum_{n=0}^{D_K-1}n\,|\varphi_n(t)|^2 .
\label{eq:C}
\end{equation}

We take as reference state the infinite-temperature TFD state,
\begin{equation}
\ket{\mathrm{TFD}}=\frac{1}{\sqrt D}\sum_{j=1}^D\ket{E_j}_L\otimes\ket{\bar E_j}_R ,
\label{eq:TFD}
\end{equation}
where $\ket{\bar E_j}$ is the complex conjugate of $\ket{E_j}$ in the occupation basis, and evolve it with $H_L$. The survival amplitude is $\bra{\mathrm{TFD}}e^{-iH_Lt}\ket{\mathrm{TFD}}=D^{-1}\sum_je^{-iE_jt}$, so the survival probability is the SFF divided by $D^2$~\cite{Balasubramanian:2022tpr,Caputa:2024vrn}. Because the TFD state has equal weight on all eigenstates, the Krylov chain depends only on the spectral measure $\mu(E)=D^{-1}\sum_j\delta(E-E_j)$: the Krylov vectors are $p_n(H_L)\ket{\mathrm{TFD}}$, where the $p_n$ are the orthonormal polynomials of $\mu$, and $a_n,b_n$ are their recurrence coefficients~\cite{Erdmenger:2026iga}. The eigenvectors of $H$ play no role, and the recursion is equivalent to running eq.~\eqref{eq:lanczos} on the diagonal matrix $\mathrm{diag}(E_1,\dots,E_D)$ with a uniform starting vector; we perform it with full reorthogonalization~\cite{Nandy:2024evd,Rabinovici:2025otw,Jeong:2026gdc}. Two properties follow at once. First, $b_1^2=\sigma_E^2$, the variance of the spectrum, so $C(t)=\sigma_E^2t^2+O(t^4)$. Second, for a non-degenerate spectrum $D_K=D$, and the long-time average of $C(t)$ is
\begin{equation}
\overline{C}=\frac1D\sum_{n=0}^{D-1}n=\frac{D-1}{2},
\label{eq:plateau}
\end{equation}
independently of the Hamiltonian. The plateau carries no information; the information lies in how it is approached, which we quantify by the overshoot (i.e., a characteristic peak structure)
\begin{equation}
\Delta C = \max_tC(t)-\frac{D-1}{2}.
\label{eq:overshoot}
\end{equation}

The unfolded Krylov complexity~\cite{Erdmenger:2026iga,Caputa:2026hvh,Basu:2026gvl} is obtained by replacing $E_j$ with the unfolded levels $\varepsilon_j$ of the full spectrum. Its spectral measure is approximately flat, so differences between unfolded complexities reflect level correlations rather than the smooth density of states; it is a spectral diagnostic that places different spectra on a common footing.

As benchmarks we use GUE and GOE random matrices of the same dimension. Their raw spectra have semicircular densities and give $\Delta C/D\simeq0.11$ (GUE) and $0.07$ (GOE); after unfolding, $\Delta C/D\simeq0.13$ and $0.081$.

\begin{figure*}[tbp]
\centering
{\small\itshape Complex couplings}\\[1pt]
\begin{subfigure}[t]{0.43\textwidth}\centering
\includegraphics[width=\linewidth]{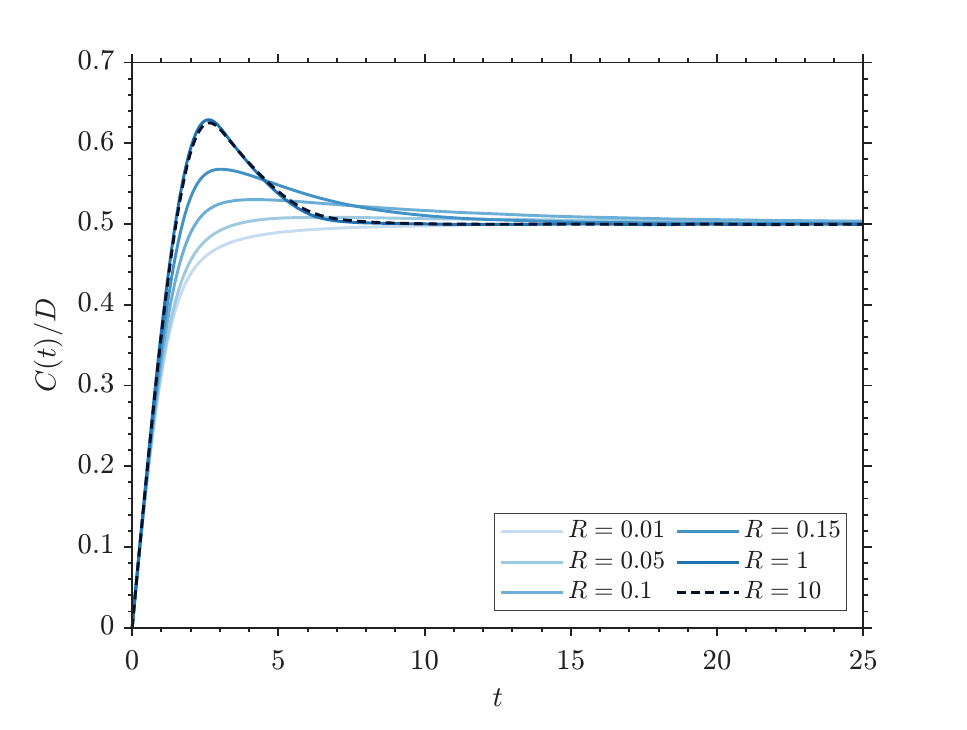}\caption{}\label{fig:unf:Cgue}
\end{subfigure}
\begin{subfigure}[t]{0.43\textwidth}\centering
\includegraphics[width=\linewidth]{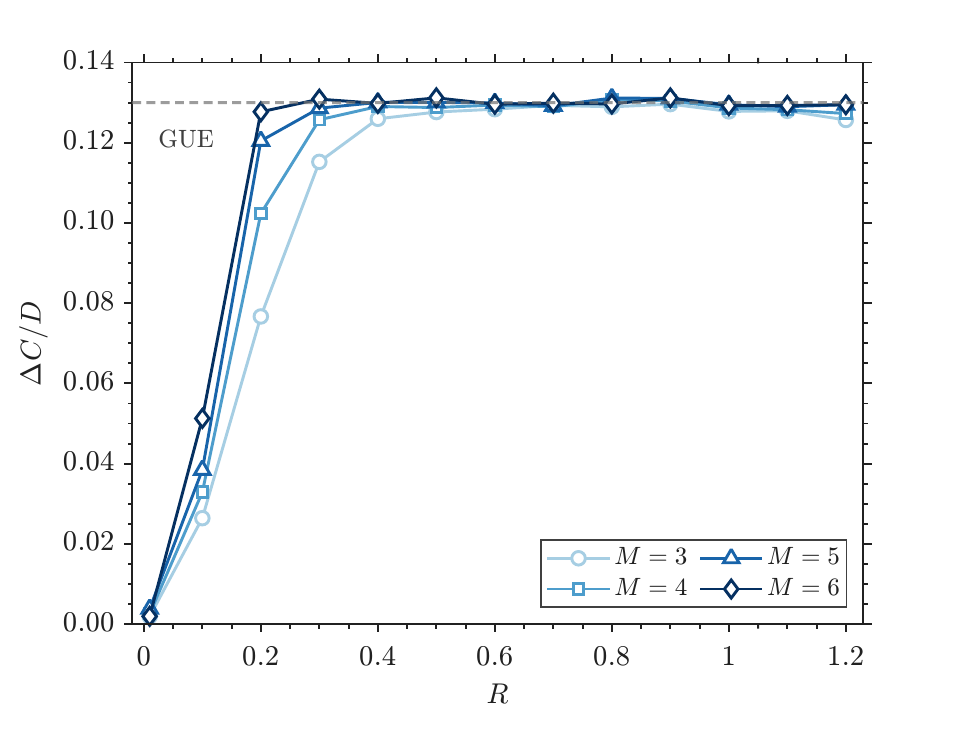}\caption{}\label{fig:unf:dCgue}
\end{subfigure}\\[4pt]

{\small\itshape Real couplings}\\[1pt]
\begin{subfigure}[t]{0.43\textwidth}\centering
\includegraphics[width=\linewidth]{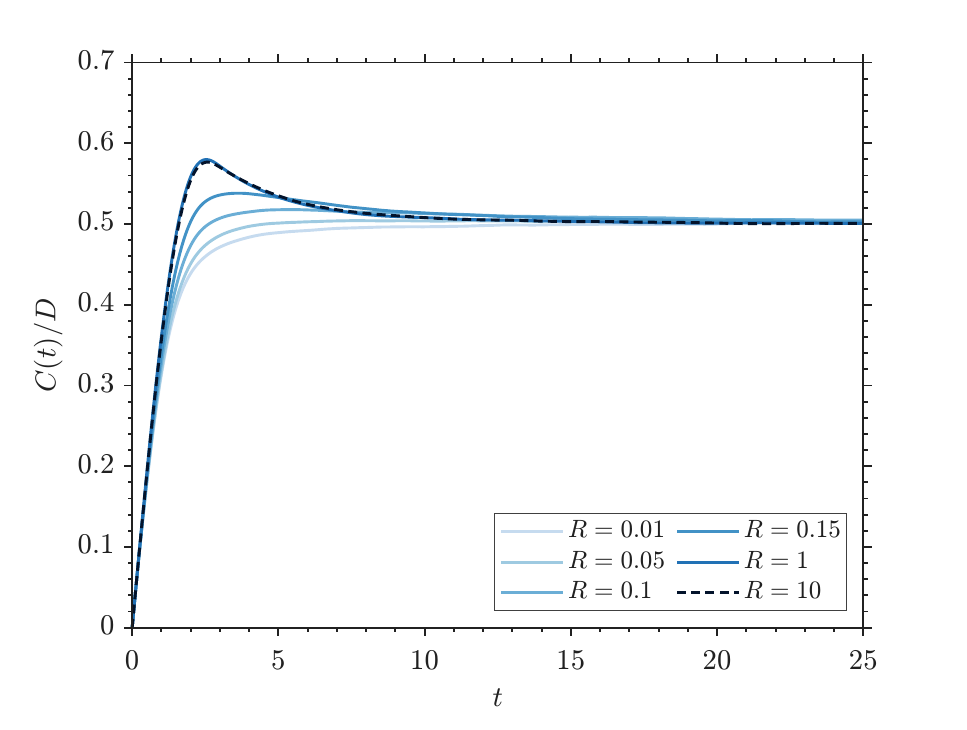}\caption{}\label{fig:unf:Cgoe}
\end{subfigure}
\begin{subfigure}[t]{0.43\textwidth}\centering
\includegraphics[width=\linewidth]{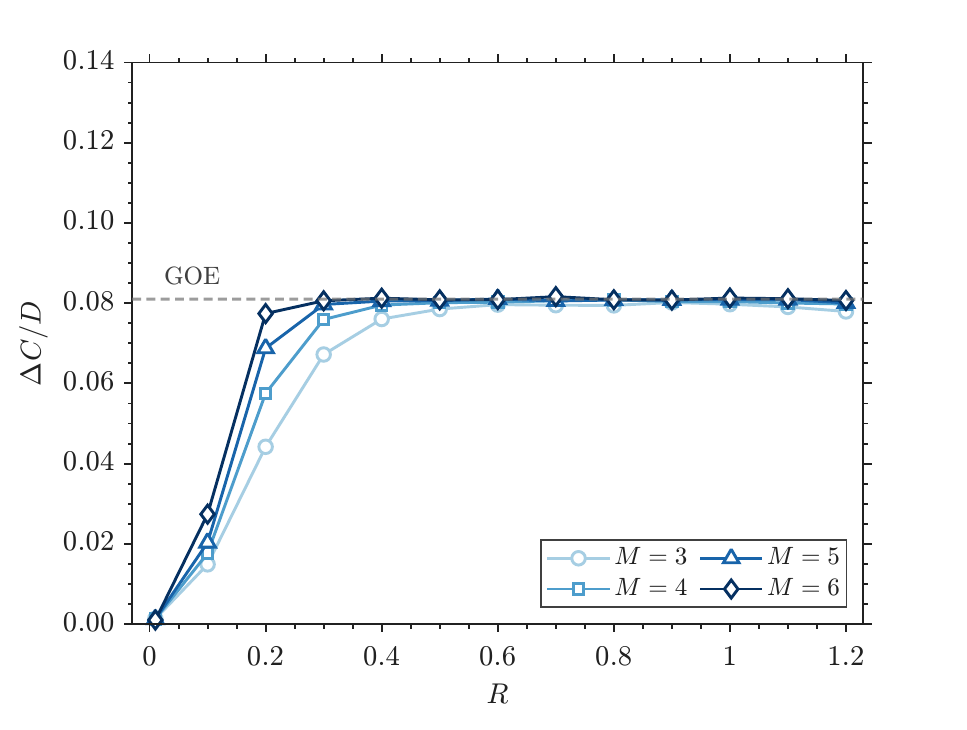}\caption{}\label{fig:unf:dCgoe}
\end{subfigure}
\caption{Krylov complexity of the TFD state computed from the unfolded spectrum when $\NB=1$. (a, c): Krylov complexity for $N=8$, $M=4$ and $500$ realizations. (b, d): The size of the peak, eq.~\eqref{eq:overshoot}, for $N=8$ and $M=3,4,5$, and $6$ ($5000$, $2000$, $500$ and $2000$ realizations, respectively). Dashed lines: unfolded GUE ($0.13$) and GOE ($0.081$) values.}
\label{fig:unfolded}
\end{figure*}

%
\section{Onset of random-matrix statistics in the spectrum of YSYK}\label{sec:spectral}

Figures~\ref{fig:Ps} and~\ref{fig:SFF} show the unfolded level spacing distribution and the unfolded SFF at three values of $R$, for complex and real couplings.

At $R=0.01$ the spacing distribution coincides with the Poisson distribution for both kinds of couplings, and the SFF reaches its plateau right after the initial transient without a correlation hole. This is what the block structure of eq.~\eqref{eq:blocks} predicts: the spectrum is a superposition of free-fermion spectra, and the random-matrix statistics of the single-particle levels inside each block leave no trace in the many-body statistics.

At $R=1$ the spacing distribution follows the Wigner surmise of the class selected by the couplings, with quadratic repulsion for complex couplings and linear repulsion for real ones. The SFF drops more than an order of magnitude below the plateau and returns to it along a ramp that ends at $t\approx t_H$. In this regime the unfolded many-body spectrum behaves like the appropriate Gaussian ensemble.

$R=0.1$ lies inside the crossover. The weight of $P(s)$ at small $s$ is reduced but remains finite, the maximum sits near $s\approx0.4$, and the tail is close to the Poisson one. The SFF has a shallow correlation hole and recovers only near $t_H$. Both observations indicate that levels are correlated only over a few mean spacings. This fits the picture of section~\ref{sec:smallR}: the oscillator term hybridizes levels of different static-boson blocks only when they are closer than the tunnelling matrix element, while levels further apart remain effectively independent. The energy range of the correlations is a few level spacings at $R=0.1$ and grows with $R$; as it grows, the correlation hole deepens and the ramp starts earlier, as seen at $R=1$.

The $R$ dependence is the same for real and complex couplings. The boson frequency controls the departure from the free-fermion regime, while the antiunitary symmetry of section~\ref{sec:symmetry} fixes the class once correlations have developed. The three values of $R$ shown here bracket the crossover; the overshoot of the Krylov complexity (given in the next section), which we have computed on a finer grid in $R$, locates it more precisely.

%
\section{Krylov complexity across the crossover}\label{sec:krylov}

\subsection{Unfolded Krylov complexity}\label{sec:unfolded}

Figure~\ref{fig:unfolded} shows the Krylov complexity computed from the unfolded spectrum, where $D_K= D$. At $R=0.01$ it rises monotonically to the plateau, as for uncorrelated levels. In figures~\ref{fig:unfolded}(a) and (c),  as $R$ increases, a peak develops at $t\approx2.5$--$3$, somewhat below $t_H/2$, and the curves for $R=1$ and $R=10$ can hardly be distinguished. The size of the peak rises from nearly zero at $R=0.01$ to the unfolded GUE value $0.13$ for complex couplings and to the GOE value $0.081$ for real couplings, and stays there up to the largest $R$ shown: see figures~\ref{fig:unfolded}(b) and (d). For $M=6$ the rise is essentially complete at $R\approx0.3$; for $M=3$ it extends to $R\approx0.4$--$0.5$. Together with the spectral data of section~\ref{sec:spectral}, this places the onset of random-matrix statistics around $R\approx0.3$.

\begin{figure*}[t]
\centering
\begin{subfigure}[t]{0.43\textwidth}\centering
\includegraphics[width=\linewidth]
{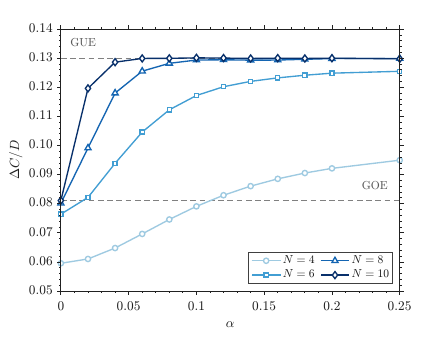}\caption{}\label{fig:alpha:dC}
\end{subfigure}
\begin{subfigure}[t]{0.43\textwidth}\centering
\includegraphics[width=\linewidth]{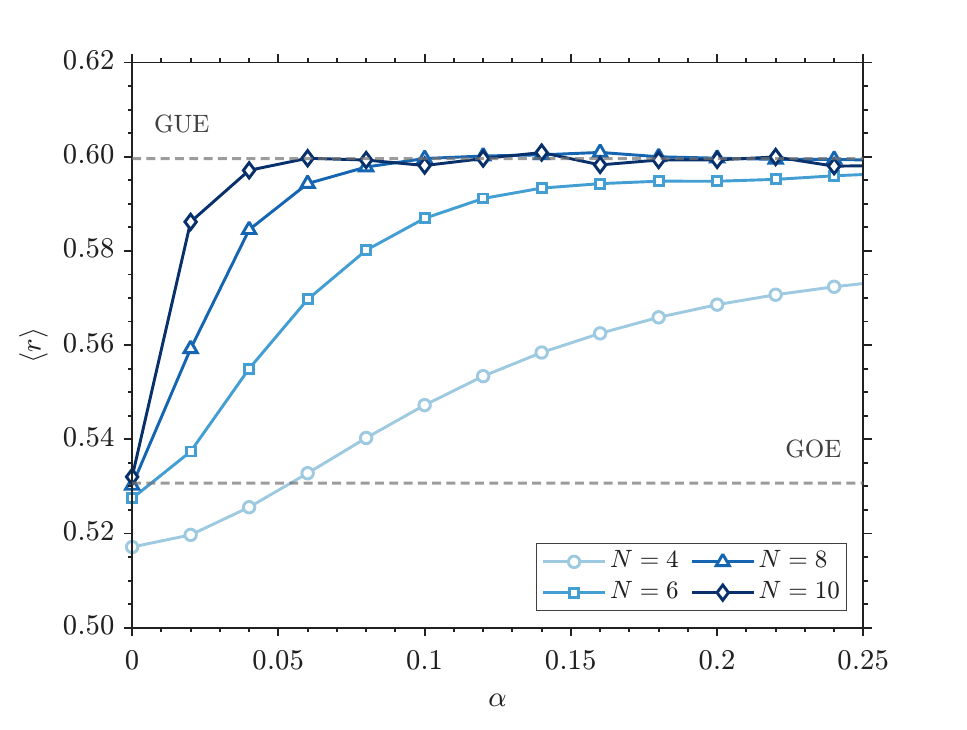}\caption{}\label{fig:alpha:r}
\end{subfigure}
\caption{Crossover from real (GOE, $\alpha=0$) to complex (GUE, $\alpha=1$) couplings, eq.~\eqref{eq:alpha}, at $R=1$ and $M=4$ for $N=4,6,8,10$ ($5000$, $2000$, $500$ and $2000$ realizations). (a) Size of the peak of the unfolded Krylov complexity, $\Delta C/D$, versus $\alpha$; dashed lines: unfolded GOE ($0.081$) and GUE ($0.13$) values. (b) Mean gap ratio $\langle r\rangle$ versus $\alpha$; dashed lines: GOE ($0.5307$) and GUE ($0.5996$) values.}
\label{fig:alpha}
\end{figure*}

The shift of the onset to smaller $R$ with increasing $M$ agrees with both limits of section~\ref{sec:model}: at small $R$ each boson flip couples a block to $M$ others, while at large $R$ the rank of the induced interaction grows with $M$. Once the size of peak has saturated, it no longer depends on $M$. The random-matrix value is reached already at $M=3$ if $R$ is large enough, in line with the observation that rank $M\geq2$ suffices for chaos and that Gaussian couplings are not needed.

It may seem surprising that the band structure at large $R$ does not affect the unfolded peak. Levels in different bands are separated by gaps much larger than the intraband spacing and are essentially uncorrelated. Unfolding maps every band onto an interval with unit mean spacing, and for sets of levels that are disjoint in energy and mutually independent, the connected SFF is the sum of their connected SFFs. After unfolding, each band has the same Heisenberg time, so a spectrum composed of several independent GUE/GOE bands has the same ramp as a single GUE/GOE spectrum of the same total size, apart from times of the order of the inverse unfolded band width. At large $R$ the unfolded peak measures correlations within the bands. Its agreement with the GUE/GOE value does not require, and does not imply, correlations between bands.

\subsection{From orthogonal to unitary statistics}\label{sec:alpha}

In order to further explore the crossover of symmetry class within the chaotic regime, we interpolate between real and complex couplings,
\begin{equation}
g_{ij,k}(\alpha)=\frac{S_{ij,k}+i\alpha A_{ij,k}}{\sqrt{1+\alpha^2}},\qquad 0\leq\alpha\leq1,
\label{eq:alpha}
\end{equation}
where $S_{ij,k}=S_{ji,k}$ and $A_{ij,k}=-A_{ji,k}$ are independent real Gaussian variables with zero mean and variance $g^2$ for $i\neq j$, and the diagonal $S_{ii,k}$ have variance $2g^2$. Then $\avg{|g_{ij,k}|^2}=g^2$ for $i\neq j$ at every $\alpha$, and $\avg{g_{ii,k}^2}=2g^2/(1+\alpha^2)$. For each $k$ the matrix $g_k(\alpha)$ belongs to the GOE at $\alpha=0$ and to the GUE at $\alpha\neq0$, where any $\alpha\neq0$ breaks the symmetry $T=K$ of section~\ref{sec:symmetry}: the GUE result in the previous subsection corresponds to the case of $\alpha = 1$.

Figures~\ref{fig:alpha}(a,b) show the Krylov complexity's peak size and the mean gap ratio at $R=1$ for various values of $N$. As $\alpha$ increases, $\Delta C/D$ from $\simeq0.08$ to $\simeq0.13$ and $\avg r$ moves from $\simeq0.53$ to $\simeq0.60$, showing the same $\alpha$ dependence between our  dynamical Krylov-based probe and statistical probe; the crossover occurs at smaller $\alpha$ in larger systems. The Krylov complexity thus tracks the symmetry class inside the chaotic regime. The appearance of the peak marks the onset of correlations, and its height has to be read against the benchmark of the relevant symmetry class.

The two observables need not behave identically, since $\avg r$ probes adjacent spacings while $\Delta C$ depends on the whole spectral measure through the Lanczos coefficients. Their agreement shows that the symmetry crossover affects the spectrum on the scales probed by both.

In the Gaussian ensemble interpolating between GOE and GUE~\cite{Pandey:1982br}, the crossover is controlled by the ratio of the typical matrix element of the antisymmetric part between neighbouring eigenstates to the mean level spacing. For a chaotic many-body spectrum of width $\sigma_E$ the matrix element scales as $\alpha \, \sigma_E/\sqrt D$, since the norm of the perturbation is spread over $D^2$ matrix elements, and the spacing as $\sigma_E/D$. The crossover should therefore occur at $\alpha_*\propto D^{-1/2}$. For $N=4,6,8,10$ we have $D=96,320,1120,4032$. The value of $\alpha$ at which $\avg r$ reaches the midpoint between its GOE and GUE values is approximated with $\propto D^{-1/2}$.

\begin{figure*}[tbp]
\centering
\begin{subfigure}[t]{0.43\textwidth}\centering
\includegraphics[width=\linewidth]{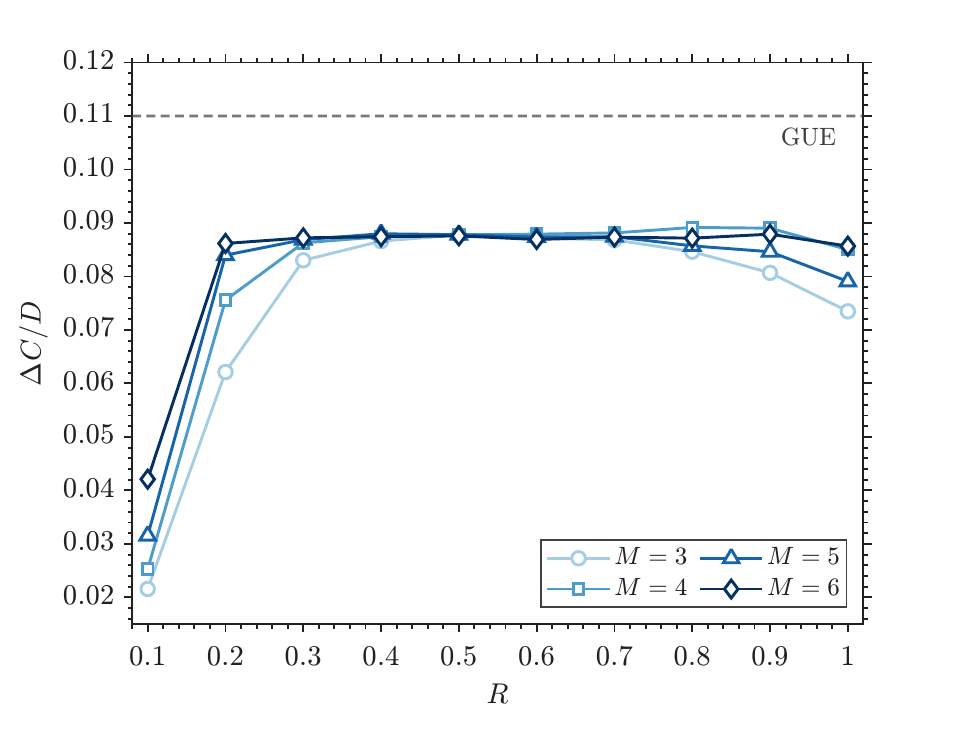}\caption{}\label{fig:raw:dC}
\end{subfigure}
\begin{subfigure}[t]{0.43\textwidth}\centering
\includegraphics[width=\linewidth]{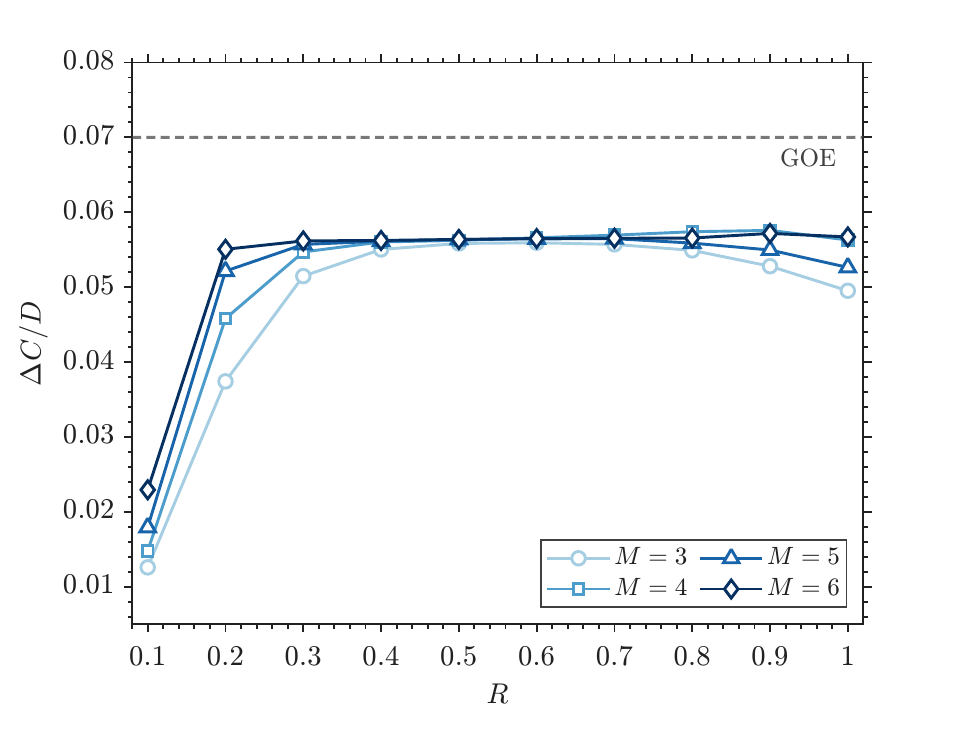}\caption{}\label{fig:app:dCgoe}
\end{subfigure}
\caption{Size of the peak of the Krylov complexity computed from the raw spectrum, $\Delta C/D$, versus $R$ for $N=8$, $\NB=1$ and $M=3,4,5,6$, with complex (a) and real (b) couplings ($2000$, $1000$, $200$ and $100$ realizations). Dashed lines: raw GUE ($0.11$) and GOE ($0.07$) values.}
\label{fig:raw}
\end{figure*}

\begin{figure*}[tbp]
\centering
\includegraphics[scale=1.0]{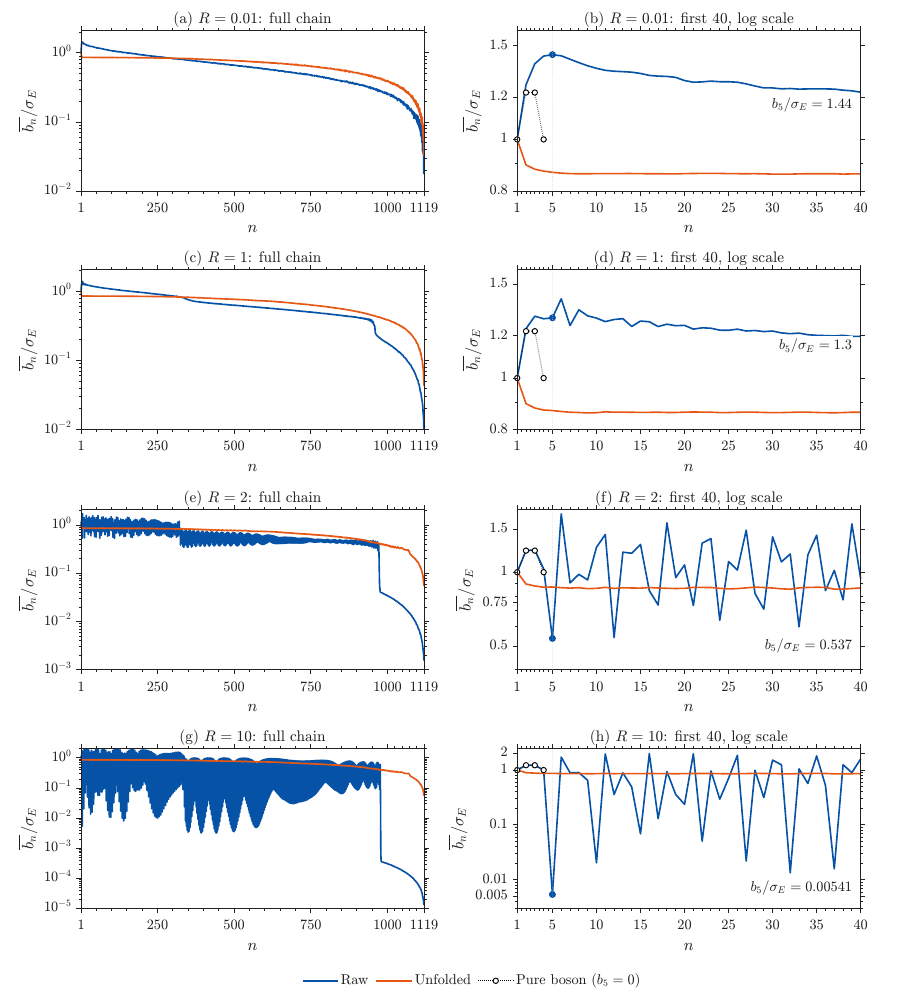}
\caption{Disorder-averaged Lanczos coefficients $b_n/\sigma_E$ for complex couplings, $N=8$, $M=4$, $\NB=1$, $100$ realizations, at $R=0.01$, $1$, $2$ and $10$. Left: full chain; right: first $40$ coefficients, on a logarithmic scale. Blue: raw spectrum; orange: unfolded spectrum. Open circles: the pure-boson chain, eq.~\eqref{eq:krawtchouk}, which terminates at $b_{M+1}=0$ (not shown on the logarithmic axes); the filled circle marks the first weak link, $b_{M+1}=b_5$, whose size relative to $\sigma_E$ controls the separation of the boson-number bands of section~\ref{sec:largeR}.}
\label{fig:lanczos}
\end{figure*}

\subsection{Lanczos coefficients and Krylov complexity from the raw spectrum}
\label{sec:raw}

We now turn to the Krylov complexity computed from the raw spectrum, which describes the evolution generated by the physical Hamiltonian and retains its density of states. Figure~\ref{fig:raw} shows the size of its peak. The peak sets in over the same range of $R$ as its unfolded counterpart (figure~\ref{fig:unfolded}) and saturates for $R\approx0.3$--$0.8$ at $\Delta C/D\simeq0.087$ for complex and $\simeq0.056$ for real couplings, below the raw GUE and GOE values $0.11$ and $0.07$; for $M=3$ and $5$ it decreases slightly as $R$ approaches one. The two ratios, $0.087/0.11\simeq0.79$ and $0.056/0.07\simeq0.80$, are the same, while the unfolded peaks agree with the ensembles in both classes. Since unfolding removes only the smooth part of the spectrum and keeps the level correlations, the deficit must come from the smooth density of states, a property to which the symmetry class is largely irrelevant; the equality of the two ratios is consistent with this. The Lanczos coefficients in figure~\ref{fig:lanczos} make the origin of the deficit explicit and show how the band structure enters at larger $R$.

Since $b_1=\sigma_E$, the normalized coefficients $b_n/\sigma_E$ start at one, and the subsequent ones encode the shape of the spectral density through its higher moments. The raw coefficients behave differently from the unfolded one. At $R=0.01$ they rise above one before decreasing slowly, and $b_2/\sigma_E\simeq1.25$. At $R=1$ the kurtosis is similar, $b_2/\sigma_E\simeq1.22$, although the density of states already shows the lobes of figure~\ref{fig:DOS}(c). At $R=2$ and $R=10$ the first four raw coefficients approach those of the pure-boson chain, while $b_5$ drops sharply. At $g=0$ ($R\rightarrow\infty$) the spectrum consists of the $M+1$ levels $\omega_0(\ell+M/2)$ with binomial weights $2^{-M}\binom{M}{\ell}$ (for $\NB=1$), whose recurrence coefficients are those of the Krawtchouk polynomials,
\begin{equation}
a_n=M\omega_0,\qquad b_n=\frac{\omega_0}{2}\sqrt{n(M-n+1)}\quad(1\leq n\leq M).
\label{eq:krawtchouk}
\end{equation}
With the variance of the spectrum $\sigma_E=\omega_0\sqrt M/2$ this gives $b_n/\sigma_E=(1,\sqrt{3/2},\sqrt{3/2},1,0)$ for $M=4$ (open circles in figure~\ref{fig:lanczos}): the pure-boson chain terminates after $M+1=5$ sites. For $g\neq0$ each level broadens into a band of width $w\propto g^2/\omega_0^2$, eq.~\eqref{eq:SW}, and $b_{M+1}$ no longer vanishes. 

Beyond $n=M+1$ the raw sequence repeats approximately with period $M+1$, with a small coefficient roughly every five steps (figure~\ref{fig:lanczos}(f,h)). The chain has turned into a sequence of short blocks joined by weak links: within a block the state moves between the $M+1$ bands, and each weak link is controlled by the structure inside the bands.

At large $R$ the raw complexity's approach to the plateau is bottlenecked by the intraband dynamics: $b_1=\sigma_E$ is still set by the coarse separation between the bands, but exploring beyond the first block of Krylov space requires the weak links $b_{M+1},b_{2(M+1)},\dots$, and their intraband origin sets the rate at which the state spreads over the rest of the chain. Unfolding closes the gaps between the bands and removes this block structure (orange curves), consistent with the coincidence of the unfolded complexities at $R=1$ and $R=10$ in figure~\ref{fig:unfolded}. The chain is never truncated: $b_{M+1}$ is small but finite, and the long-time average of $C(t)$ remains the plateau~\eqref{eq:plateau}. The raw deficit is therefore not a sign of weaker chaos. At moderate $R$ it reflects the non-semicircular density of states, and at large $R$ the band structure further separates the raw dynamics from that of a random matrix. A raw many-body complexity should thus not be compared with semicircular random-matrix benchmarks; unfolding provides the common footing.

%
\section{Discussion}\label{sec:discussion}

The YSYK model reaches many-body chaos through a boson that is static at small $R$ and virtual at large $R$. The control parameter $R=\omega_0/g^{2/3}$ compares the boson frequency with the fermionic energy scale $E_Y=g/\sqrt{\omega_0}$ set by the Yukawa coupling, which is why a stronger coupling at fixed $\omega_0$ makes the spectrum less chaotic. For $R\ll1$ the boson coordinates are conserved and the fermions are free in each of the $(\NB+1)^M$ boson configurations, so the many-body levels are uncorrelated. For $R\gg1$ the bosons generate, within each band of fixed boson number, a four-fermion interaction of rank $M$, which is generically chaotic for $M\geq2$, and approaches a complex SYK$_4$ interaction at large $M$. The random-matrix class is fixed by the reality of the couplings: complex couplings leave no antiunitary symmetry and give the GUE, real couplings give the GOE.

Between these limits, the spacing distribution, the unfolded SFF and the unfolded Krylov complexity of the TFD state agree that random-matrix statistics set in around $R\approx0.3$. Inside the crossover, at $R=0.1$, levels are correlated only over a few mean spacings, consistent with the oscillator term hybridizing levels of different static-boson blocks only when they are nearly degenerate. More boson modes shift the onset to smaller $R$, as explored from both limits, and once the peak of the Krylov complexity has saturated it no longer depends on $M$: the random-matrix value is reached already for $M=3$, and Gaussian effective couplings are not needed.

For Krylov complexity the lessons are twofold. After unfolding, the peak appears when the levels become correlated and reaches the value of the relevant Gaussian ensemble, even when the raw spectrum is split into boson-number bands, because it then measures the correlations within each band. Along the interpolation between real and complex couplings it follows the mean gap ratio, with a crossover that moves to smaller $\alpha$ in larger systems, consistent with the estimate $\alpha_*\propto D^{-1/2}$. The unfolded peak is thus a reliable indicator of both the onset of chaos and the symmetry class, provided its height is read against the benchmark of the relevant random-matrix class. 

Without unfolding, the peak also reflects the density of states. Where the density is smooth but far from a semicircle, the raw peak saturates at about $80\%$ of the raw GUE and GOE values, the same fraction in both classes. At larger $R$ the boson-number bands turn the Krylov chain into blocks of $M+1$ sites joined by weak links whose size is set by the intraband width; these links do not truncate the chain but slow the spreading over it. In many-body systems whose density of states is far from a semicircle, or which are organized into bands by approximately conserved quantities, the spectrum should therefore be unfolded before the peak is used as an order parameter for chaos.

Our results have clear limitations. The Hilbert spaces are small ($D\leq4480$ in the frequency scans), and the crossover values of $R$ may shift with $N$. The bosons are truncated to $\NB=1$. This is adequate at large $R$, where only single virtual quanta matter, but at small $R$ the untruncated bosons would be displaced far from their vacuum, and the static-boson regime of the full model need not coincide quantitatively with that of the truncated one. All results are at infinite temperature, whereas the large-$N$ analysis of ref.~\cite{Davis:2022iqi} concerns low temperatures and the competition between non-Fermi-liquid and insulating phases. Extending the present diagnostics to the TFD state at finite temperature, which weights the low-energy part of the spectrum, and to larger $\NB$ would connect the two. Charge sectors away from half filling, where particle-hole conjugation no longer maps the sector onto itself, are another natural extension.

The system sizes considered here are those relevant for the cavity-QED proposal of ref.~\cite{PascualSolis:2025cuc} (see also ref.~\cite{Pelliconi:2026hoy} for dissipative effects of a lossy cavity). Our results indicate that such a simulator needs $R\gtrsim0.3$ to realize SYK-type many-body chaos, and that the effective truncation of the cavity modes deserves attention in that context.

Random-matrix spectral correlations are a prerequisite for any holographic interpretation of SYK-type models. In the SYK model, the ramp of the spectral form factor and the Krylov complexity of the thermofield-double state admit bulk interpretations in terms of the double-cone geometry~\cite{Kitaev2015Talk,Maldacena:2016hyu,Cotler:2016fpe,Kitaev:2017awl} and the length of the Einstein–Rosen bridge~\cite{Lin:2022rbf,Rabinovici:2023yex,Heller:2024ldz,Fu:2025kkh}, respectively. Whether these interpretations carry over to the YSYK model, whose low-temperature physics differs from that of SYK, remains open. Our results delineate the regime, $R\gtrsim0.3$, in which the necessary late-time chaotic structure is present, and thus where a cavity-QED platform may be used to address this question.

Finally, the spatially extended YSYK model with spatially random couplings yields linear-in-temperature resistivity~\cite{Patel:2022gdh}, one of the central puzzles of strange metals~\cite{Lee2006,Anderson2017,Lee2018,Greene2020,Varma2020,Hartnoll2022,Phillips2022}. Related results have been obtained for spatially random couplings to vector bosons~\cite{Wang:2024utm,Wang:2025oiz}, and ref.~\cite{Sin:2025sue} argued that at the quantum critical point the Yukawa form is the unique scalar coupling with this property. In that theory the boson acquires a thermal mass $\Delta(T)$~\cite{Esterlis2021,Guo2022}. By analogy with the present results, the ratio of $\Delta(T)$ to the Yukawa-induced fermionic energy scale may play the role of $R$, so that the chaotic character of the dynamics would change with temperature. Testing this requires finite-temperature diagnostics and ingredients absent in $0+1$ dimensions, such as the Fermi surface and momentum conservation~\cite{Patel:2016wdy,Singh:2026etp}; the present analysis infers no direct connection to the strange metal transport, which remains an open question.

%
\acknowledgments
We would like to thank {Hugo A. Camargo}, {Antonio Miguel Garc\'ia-Garc\'ia}, {Henning Schomerus}, {Sang-Jin Sin} and {Zhuo-Yu Xian} for valuable discussions and correspondence. 
HSJ was supported by an appointment to the JRG Program at the APCTP through the Science and Technology Promotion Fund and Lottery Fund of the Korean Government. HSJ was also supported by the Korean Local Governments -- Gyeongsangbuk-do Province and Pohang City.
YLW is supported by an appointment to the Young Scientist Training Programme at the APCTP through the Science and Technology Promotion Fund and Lottery Fund of the Korean Government.

%
\appendix
\section{Path-Integral Interpretation of the SYK$_2$-like to SYK$_4$-like Crossover}\label{app:pi}
\begin{figure}[t]
   \centering
    \includegraphics[width=0.50\linewidth]{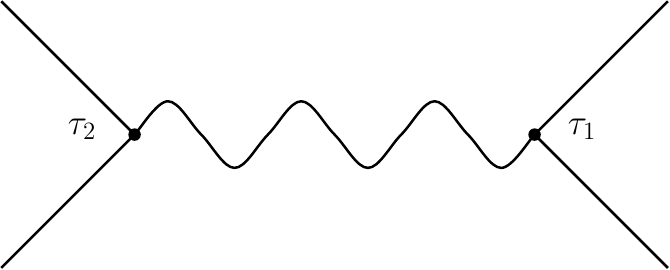}
    \caption{Interaction between two fermion pairs (straight lines) at $\tau_1$ and $\tau_2$ via the exchange of a boson (wavy line). In coordinate space, the temperal correlation of the force is $1/\omega_0$. }\label{fig:scattering}
\end{figure}

The SYK$_{q=2,4}$-like behaviour of a YSYK system can be understood in a heuristic way. Yukawa-type theories enable a non-local interaction between fermions pairs, $\psi^\dagger\psi$, at imaginary times $\tau_1$ and $\tau_2$, and the interaction force is given by the exchange of a boson $\phi$, as is demonstrated by figure~\ref{fig:scattering}. The
characteristic range (temporal correlation) of this interaction is $1/\omega_0$ \cite{Weinberg_1995}. When $\omega_0\gg 1$, the correlation length is so small that the interaction is almost local, which is effectively a four-fermion vertex. On the other hand, if $\omega_0\ll 1$, the temporal correlation length between fermions pairs is long, so the exchange of bosons is slow. From the perspective of fermions, the scalar field is effectively a background rather than a dynamical object, \textit{i.e.} bosons cannot efficiently mediate dynamical interactions between fermion pairs. In this case, the interaction reduces to two-fermion-like.\\

However, the interpretation above can only naively tell how $\omega_0$ affect the system, which cannot explain how the effective coupling constant becomes a Gaussian. In addition, numerical observations also show large-$M$ approximate SYK$_4$ better at large $\omega_0/g^{2/3}$, and the scattering picture is unable to explain such a dependence on bosonic flavour number either. 
It may also still seem counterintuitive that a stronger interaction (larger $g$) leads to less chaotic behaviour. 
The similarity between YSYK and SYK$_{q=2,4}$, as well as why $\omega_0/g^{2/3}$, rather than $\omega_0$ alone, serves as the relevant controlling parameter, can be understood from the perspective of the path-integral formalism, as we now illustrate. \\

In terms of {Euclidean action}, the theory is given by 
\begin{eqnarray}
    S\equiv [\psi^\dagger,\psi]+S[\phi]+S_{\text{YSYK}},
\end{eqnarray}
where
\begin{eqnarray}
    &&S[\psi^\dagger,\psi]=\int d\tau \sum_{i=1}^N \psi_i^\dagger(\tau) (\partial_\tau-\mu) \psi(\tau),\\
    &&S[\phi]=\frac{1}{2}\int d\tau \sum_{i=1}^N \phi_i(\tau)\left[-\partial_\tau^2 +\omega_0^2\right]\phi_{i}(\tau),\\
    &&S_{\text{YSYK}}=\int d\tau\sum_{i,j}^N\sum_l^M \frac{g_{ijl}}{\sqrt{NM}} \psi^\dagger_{i}(\tau)\psi_{j}(\tau)\phi_l(\tau).
\end{eqnarray}

The partition function is given by
\begin{eqnarray}
    \mathcal{Z}&=&\int \mathcal{D}[\psi^\dagger,\psi]\mathcal{D}[\phi]e^{-S}\nn
    &=&\int \mathcal{D}[\psi^\dagger,\psi]\exp\left\{-\int d\tau \sum_{i=1}^N \psi_i^\dagger(\tau) (\partial_\tau-\mu) \psi(\tau)\right\}\nn
    &&\times\int\mathcal{D}[\phi]\exp\left\{-\frac{1}{2}\int d\tau \sum_{i=1}^N \phi_i(\tau)\left[-\partial_\tau^2 +\omega_0^2\right]\phi_{i}(\tau)\right.\nn
    &&\left.-\int d\tau\sum_{i,j}^N\sum_l^M \frac{g_{ijl}}{\sqrt{NM}} \psi^\dagger_{i}(\tau)\psi_{j}(\tau)\phi_l(\tau)\right\}\nn
    &=&\int \mathcal{D}[\psi^\dagger,\psi]\exp\left\{-\int d\tau \sum_{i=1}^N \psi_i^\dagger(\tau) (\partial_\tau-\mu) \psi(\tau)\right.\nn
    &&+\left.\frac{1}{2} \sum_{i,j,i',j'=1}^N\sum_{l=1}^M\frac{g_{ijl}g^*_{i'j'l}}{N M}\psi_i^{\dagger}(\tau)\psi_j(\tau)\frac{1}{-\partial_\tau^2+\omega_0^2}\psi_{j'}^\dagger(\tau)\psi_{i'}(\tau)\right\}.\nn
\end{eqnarray}
Integrating out bosonic degrees of freedom will effectively yield 
\begin{eqnarray}
    S=S[\psi^\dagger,\psi]+S_{\text{eff}},
\end{eqnarray}
where
\begin{eqnarray}\label{eq:effectiveS}
    S_{\text{eff}}&\equiv&-\frac{1}{2}\int d\tau \sum_{i,j,i',j'=1}^N\sum_{l=1}^M\frac{g_{ijl}g^*_{i'j'l}}{N M}\nn
    &&\psi_i^{\dagger}(\tau)\psi_j(\tau)\frac{1}{-\partial_\tau^2+\omega_0^2}\psi_{j'}^\dagger(\tau)\psi_{i'}(\tau).
\end{eqnarray}
We will now show how the effective action \eqref{eq:effectiveS} can lead to SYK$_4$-like and SYK$_2$-like behaviour respectively at large and small $\omega_0/g^{3/2}$. 
\\

\paragraph{SYK$_4$ limit} Let $E_f$ be the characteristic energy of the fermions. 
 Let us define
\begin{eqnarray}
    g^2_{iji'j',l}\equiv -g_{ijl}g^*_{i'j'l}.
\end{eqnarray}
If $E_f\ll \omega_0$, the effective action \eqref{eq:effectiveS} takes the form
\begin{eqnarray}\label{eq:effs1}
    S_{\text{eff}}=\int d\tau \sum_{i,j,i',j'=1}^N\frac{1}{NM}\sum_{l=1}^M\frac{g^2_{iji'j',l}}{2\omega_0^2}\psi_i^{\dagger}(\tau)\psi_j(\tau)\psi_{j'}^\dagger(\tau)\psi_{i'}(\tau),\nn
\end{eqnarray}
to the leading order.
We have obtained an interaction involving four fermions, but the coupling parameter of four-fermion interaction,
\begin{eqnarray}
    \Tilde{g}_{iji'j'}\equiv \frac{1}{M}\sum_{l=1}^M g^2_{iji'j',l},
\end{eqnarray}
generally does not obey Gaussian distribution. The next step is thus to figure out how can we find a four-fermion coupling governed by a Gaussian with zero mean, in order to reproduce an SYK$_4$-type interaction. \\
Note that every $g^2_{iji'j',l}$ has an non-vanishing expectation value
\begin{eqnarray}
    \langle g^2_{iji'j',l}\rangle=\langle g_{ijl}g^*_{i'j'l}\rangle=g^2\delta_{ii',jj'}\neq 0.
\end{eqnarray}
The variance of $g^2_{iji'j',l}$, denoted by $\sigma^2$, is also non-vanishing and finite. One finds
\begin{eqnarray}
    &&\langle g^*_{i_1j_1l}g_{i_2j_2l}g^*_{i_3j_3l}g_{i_4j_4l}\rangle\nn
    &=&
    \langle g^*_{i_1j_1l}g_{i_2j_2l}\rangle \langle g^*_{i_3j_3l}g_{i_4j_4l}\rangle+\langle g^*_{i_1j_1l}g_{i_4j_4l}\rangle \langle g^*_{i_3j_3l}g_{i_2j_2l}\rangle\nn
    &=&g^4\delta_{i_1i_2}\delta_{i_3i_4}\delta_{j_1j_2}\delta_{j_3j_4}+g^4\delta_{i_1i_4}\delta_{i_3i_2}\delta_{j_1j_4}\delta_{j_3j_2},
\end{eqnarray}
and 
\begin{eqnarray}
    \langle g^*_{i_1j_1l}g_{i_2j_2l}\rangle \langle g^*_{i_3j_3l}g_{i_4j_4l}\rangle=g^4 \delta_{i_1i_2}\delta_{i_3i_4}\delta_{j_1j_2}\delta_{j_3j_4}.
\end{eqnarray}
Thus,
\begin{eqnarray}
    \sigma^2&=&\langle g^*_{i_1j_1l}g_{i_2j_2l}g^*_{i_3j_3l}g_{i_4j_4l}\rangle^2-\langle g^*_{i_1j_1l}g_{i_2j_2l}\rangle \langle g^*_{i_3j_3l}g_{i_4j_4l}\rangle\nn
    &=&g^4\delta_{i_1i_4}\delta_{i_3i_2}\delta_{j_1j_4}\delta_{j_3j_2}<\infty.
\end{eqnarray}
According to the central limit theorem \cite{CLT}, the distribution of 
\begin{eqnarray}
    J_{iji'j'}\equiv \sqrt{M}\left(\Tilde{g}_{iji'j'}-g^2\delta_{ii',jj'}\right)
\end{eqnarray}
will be Gaussian with a zero mean and a variance $\sigma^2$ when we take $M\to\infty$. The effective action \eqref{eq:effectiveS} can therefore be rewritten into
\begin{eqnarray}\label{eq:efff}
    &&S_{\text{eff}}\nn
    &=&\int d\tau \sum_{i,j,i',j'=1}^N\frac{1}{N}\frac{1}{2\omega_0^2}\left[\frac{J_{iji'j'}}{\sqrt{M}}\psi_i^{\dagger}(\tau)\psi_j(\tau)\psi_{j'}^\dagger(\tau)\psi_{i'}(\tau)\right.\nn
   &&\left. +g^2\psi_i^{\dagger}(\tau)\psi_j(\tau)\psi_{j}^\dagger(\tau)\psi_{i}(\tau)\right],
\end{eqnarray}
where
\begin{eqnarray}
    \langle J_{iji'j'}\rangle=0,\quad \langle J_{iji'j'}J^*_{aba'b'}\rangle=g^4\delta_{ia}\delta_{jb}\delta_{i'a'}\delta_{j'b'}.
\end{eqnarray}
So, the term governed by $J_{iji'j'}$ in eq.~\eqref{eq:efff} becomes an SYK$_4$-type interaction. Meanwhile, the other term in eq.~\eqref{eq:efff}
\begin{eqnarray}
    \sum_{ij}\psi^\dagger_i\psi_j\psi^\dagger_j\psi_i=\sum_{i,j=1}^N n_i(1-n_j)
\end{eqnarray}
only contains the total fermion number, which is a conserved quantity, so this term is of no consequence for the purpose of quantum-chaos diagnostics. \\

Furthermore, in a $(0+1)$D system, the mass dimensions of fields and coupling constant are
\begin{eqnarray}
    &&[\psi]=0,\qquad[\phi]=-\frac{1}{2},\qquad[g]=\frac{3}{2}.
\end{eqnarray}

One can verify that $g^2_{iji'j',l}/\omega_0^2$ has mass dimension $1$, and according to eq.~\eqref{eq:effs1}, $g^2/\omega_0^2$ can be regarded as the characteristic fermionic energy scale of induced SYK$_4$ interaction. Then the condition that $E_f\ll \omega_0$ is equivalent to
\begin{eqnarray}
    E_f\sim \frac{g^2}{\omega_0^2}\ll\omega_0,
\end{eqnarray}
which leads to
\begin{eqnarray}
    \frac{\omega_0}{g^{2/3}}\gg 1.
\end{eqnarray}
and $\omega_0/g^{2/3}$ is dimensionless. As the true controlling parameter, $\omega_0/g^{2/3}$ reflects the competition between fermionic and bosonic energies, rather than bosonic mass and the coupling strengh. In summary, {large $\omega_0/g^{2/3}$, together with large $M$, can give rise to an SYK$_4$-like chaotic dynamics from an YSYK system.}\\

\paragraph{SYK$_2$ limit}Now let us consider the opposite case, where ${\omega_0}/{g^{2/3}}\ll 1$. Let 
\begin{eqnarray}
    \mathcal{O}_{ij}(\tau)\equiv \psi_i^\dagger(\tau)\psi_j(\tau).
\end{eqnarray}
In frequency representation, action \eqref{eq:effectiveS} becomes
\begin{eqnarray}
    S_{\text{eff}}=-\frac{1}{2}\int \frac{d\nu}{2\pi}\sum_{i,j,i',j'=1}^N\sum_{l=1}^M\frac{g_{ijl}g^*_{i'j'l}}{N M} \mathcal{O}_{ij}(\nu)\frac{1}{\nu^2+\omega_0^2}\mathcal{O}_{j'i'}(-\nu).\nn
\end{eqnarray}
Cauchy representation of Dirac delta function takes the form
\begin{eqnarray}
    \lim_{a\to0}\frac{a}{\pi(a^2+x^2)}=\delta(x).
\end{eqnarray}
Thus at small $\omega_0$,
\begin{eqnarray}
    S_{\text{eff}}&\simeq&-\frac{1}{4}\int d\nu\sum_{i,j,i',j'=1}^N\sum_{l=1}^M\frac{g_{ijl}g^*_{i'j'l}}{\omega_0 N M} \mathcal{O}_{ij}(\nu)\mathcal{O}_{j'i'}(-\nu)\delta(\nu)\nn
    &=&-\frac{1}{4}\sum_{i,j,i',j'=1}^N\sum_{l=1}^M\frac{g_{ijl}g^*_{i'j'l}}{\omega_0 N M} \mathcal{O}_{ij}(0)\mathcal{O}_{j'i'}(0)\nn
    &=&-\frac{1}{4}\int_0^\beta d\tau \int_0^\beta d\tau'\sum_{i,j,i',j'=1}^N\sum_{l=1}^M\frac{g_{ijl}g^*_{i'j'l}}{\omega_0 N M} \mathcal{O}_{ij}(\tau)\mathcal{O}_{i'j'}^\dagger(\tau').\nn
\end{eqnarray}
Hence,
\begin{eqnarray}
    &&e^{-S_{\text{eff}}}\nn
    &=&\exp\left\{\frac{1}{4}\int_0^\beta d\tau \int_0^\beta d\tau'\sum_{i,j,i',j'=1}^N\sum_{l=1}^M\frac{g_{ijl}g^*_{i'j'l}}{\omega_0 N M} \mathcal{O}_{ij}(\tau)\mathcal{O}_{i'j'}^\dagger(\tau')\right\}\nn
   &=& \int dJ \exp{-\frac{J^2}{2}-J \sum_{i,j=1}^N \sum_{l=1}^M \frac{g_{ijl}}{\sqrt{NM}}\int_0^\beta d\tau \mathcal{O}_{ij}(\tau)},
\end{eqnarray}
where $J$ is a {non-dynamical} variable {independent} of $\tau$. As a result, one finds for small $\omega_0$, \eqref{eq:effectiveS} takes the form
\begin{eqnarray}\label{eq:effs2}
     &&S_{\text{eff}}[\psi^\dagger,\psi]\nn
     &\simeq&\frac{J^2}{2}+J \sum_{i,j=1}^N \sum_{l=1}^M \frac{g_{ijl}}{\sqrt{NM}}\int d\tau \psi^\dagger_i(\tau)\psi_j(\tau),
\end{eqnarray}
which are approximately a set of an SYK$_2$ theory with various values of $l$. As $J$ has no dynamical d.o.f., it only has trivial contribution that is of no consequence.

In fact, the action \eqref{eq:effs2} merely replaces the scalar field $\phi(\tau)$ in  with $J$.
Although the Hubbard–Stratonovich transformation formally brings the action back to a form resembling the original one, the path-integral formulation makes the physical content of the limiting procedure transparent. The original dynamical field $\phi(\tau)$ is effectively reduced to a non-dynamical constant $J$. Small bosonic mass implies low frequency, so bosons are relatively ``static'' from the perspective of fermions. Therefore, every configuration of $\phi(\tau)$ can be effectively described by a constant, which enables the emergence of SYK$_2$ properties.\\

%
\bibliography{biblio}
\bibliographystyle{JHEP}

\end{document}